\documentclass[fleqn,usenatbib,useAMS]{mnras}
\usepackage{graphicx}	% Including figure files
\usepackage{amsmath}	% Advanced maths commands
\usepackage{amssymb}	% Extra maths symbols
\usepackage{multicol}        % Multi-column entries in tables
\usepackage{bm}		% Bold maths symbols, including upright Greek
\usepackage{pdflscape}	% Landscape pages
\usepackage{aas_macros}
\usepackage{ulem}
\usepackage{color}


\makeatletter
\def\ion#1#2{#1$\;${\small\rm\@Roman{#2}}\relax}
\newcommand{\phn}{\phantom{0}}

\newcommand{\myemail}{t.hayashi@nao.ac.jp}
\makeatother
\usepackage[T1]{fontenc}
\usepackage{ae,aecompl}
\usepackage{newtxtext,newtxmath}
\title[Radio properties of nearby ULIRGs with buried AGN signatures]{
	Radio properties of ten nearby ultraluminous infrared galaxies with 
	signatures of luminous buried active galactic nuclei
}
\author[T.~J.~Hayashi et al.]{
	Takayuki J. Hayashi$^{1,2}$
	\thanks{Contact-mail: \href{mailt:\myemail}{\myemail}}, 
	Yoshiaki Hagiwara$^{3}$, 
	and Masatoshi Imanishi$^{1}$\\
    $^{1}$National Astronomical Observatory of Japan, 2-21-1 Osawa, Mitaka, Tokyo 181-8588, Japan\\
	$^{2}$Azabu Junior and Senior High School, 2-3-29 Motoazabu, Minato-ku, Tokyo 106-0046, Japan\\
	$^{3}$Toyo University, Hakusan 5-28-20, Bunkyo-ku, Tokyo 112-8606, Japan
}
\date{\today}
\pubyear{2020}
\begin{document}
\label{firstpage}
\pagerange{\pageref{firstpage}--\pageref{lastpage}}
\maketitle

%%%%%%%%%%%%%%%%%%%%%%%%%%%%%%%%%%%%%%%%%%%%%%%%%%
% Abstract of the paper
\begin{abstract}
We present the results of our multifrequency observations for 10 ultraluminous infrared galaxies (ULIRGs) made by the Karl G. Jansky Very Large Array at 1.4, 5.5, 9.0, and 14.0\,GHz. 
Our sample is selected from ULIRGs whose active galactic nuclei (AGNs) are not found at optical wavelengths ($\sim$70\% of the entire ULIRGs), but whose presence is suggested by mid-infrared or submillimetre observations ($>50$\% of the non-AGN ULIRGs at optical wavelengths). 
The statistical properties of the targets are similar to those of the entire ULIRG sample, which implies that ULIRGs have common radiative processes regardless of the presence of optical AGNs, and thus AGNs might equally contribute to the radio emission of every ULIRG.
Although their spectra are mainly explained by starbursts and/or merger activity, some individual sources suggest contributions from AGNs.
IRAS\,00188$-$0856, whose optical morphology is not disturbed, shows a large nonthermal fraction and a spectral break at high frequency, which can be explained by synchrotron ageing of nonthermal plasma emitted from AGNs.
In addition, we find 100-kpc scale extended emission associated with IRAS\,01004$-$2237.
The two-sided morphology and absence of extended X-ray emission suggest that this system is not induced by a merger in a cluster but originates from AGN activity.
\end{abstract}

\begin{keywords}
galaxies: active --- galaxies: nuclei --- radio continuum: galaxies --- techniques: interferometric
\end{keywords}
%%%%%%%%%%%%%%%%%%%%%%%%%%%%%%%%%%%%%%%%%%%%%%%%%%

%%%%%%%%%%%%%%%%% BODY OF PAPER %%%%%%%%%%%%%%%%%%
% -----------------------------------------------------------------------------------
\section{Introduction}
% -----------------------------------------------------------------------------------
Galaxies with large infrared luminosities greater than $10^{12}L_{\odot}$ are known as ultraluminous infrared galaxies \citep[ULIRGs;][]{1996ARA&A..34..749S}, whose powerful energy sources are hidden behind infrared-emitting dust heated by energetic radiation from starbursts and/or active galactic nuclei (AGNs).
Currently, there is a debate as to whether starbursts or AGNs are more energetically important.
This is closely linked to understanding the history of star formation and the growth of supermassive black holes in the dust-obscured galaxy population in the early Universe.
Unlike optically identifiable AGNs surrounded by torus-shaped dust \citep[optical Seyferts;][]{1987ApJS...63..295V,2001ApJS..132...37K,2003MNRAS.346.1055K}, ULIRGs are systems of major mergers and contain large amounts of concentrated molecular gas and dust in their nuclei \citep{1996ARA&A..34..749S}.   
The putative compact AGNs at ULIRG's nuclei can thus be completely obscured by the gas and dust in all directions. 
Therefore, while $\sim$30\% of ULIRGs are classified as Seyfert \citep{1999ApJ...522..113V}, this situation makes optical detection of AGN signatures difficult. 
Hence, the existence of elusive buried AGNs is presumed \citep{2003MNRAS.344L..59M,2006ApJ...637..114I}.

To investigate elusive buried AGNs in ULIRG's nuclei, 
it is critical to observe the wavelengths where the effects of dust extinction are small.
Based on spectroscopy at the mid-infrared (MIR) band of 3--35\,$\mu$m, 
buried AGNs are distinguished from normal starbursts and have also been identified in more than half of ULIRGs, where no AGNs are found at optical wavelengths \citep{2004ApJS..154..178A,2007ApJ...656..148A,2007ApJS..171...72I,2008PASJ...60S.489I,2010ApJ...709..801I,2010ApJ...721.1233I,2009ApJ...694..751I,2008MNRAS.385L.130N,2009MNRAS.399.1373N,2010MNRAS.405.2505N,2009ApJS..182..628V}.
Although it is considered physically extreme
\citep{2000AJ....119..509S,2005ApJ...630..167T,2007ApJS..171...72I}, starbursts highly concentrated in the centre may show infrared spectra similar to those of buried AGNs \citep{2004A&A...414..873S}.
Thus, the energetic contribution of AGNs has not been firmly established yet.
Furthermore, recent submillimetre observations report a high flux-density ratio of HCN to HCO$^{+}$ emission, suggesting that deeply buried AGNs are present in ULRIGs even without signatures of AGNs at MIR \citep{2016AJ....152..218I,2018ApJ...856..143I,2019ApJS..241...19I}.
%,2020ApJ...902...99I}
There is a physical limitation to scrutinise ULIRG's nuclei by infrared observations alone.
Hence, we require observations at another wavelength, such as radio wave, where dust is more transparent than at infrared or submillimetre wavelengths.  

In this study, we focus on radio spectra of ULIRGs.
Radio emission from ULIRGs is a mixture of those originating from thermal and nonthermal plasma, where the former originates from star formation while the latter is also related to AGNs \citep{1991ApJ...378...65C}.
In the GHz range, although an optically thin steep spectrum is produced by nonthermal plasma in supernova remnants, it sometimes becomes peaked or inverted in AGNs if synchrotron self-absorption (SSA) owing to a strong magnetic field is present \citep[e.g.][]{2005A&A...435..839T,2008A&A...487..885O}.
Moreover, the large bulk Lorentz factor of a collimated jet from AGNs can cause strong variability \citep{2001ApJ...561..676L}.
Thus, the presence of AGNs can be confirmed by studying the spectral shape and flux-density variation.
In addition, while typical ULIRGs show steep spectra with flattening at low and high frequencies caused by free-free absorption (FFA) and free-free emission (FFE) due to thermal plasma, respectively  \citep{1991ApJ...378...65C,2008A&A...477...95C,2010MNRAS.405..887C,2011ApJ...739L..25L,2018MNRAS.474..779G,2013ApJ...777...58M}, 
some sources suggest steepening at high frequency.
While \citet{2010MNRAS.405..887C} and \citet{2013ApJ...777...58M}  consider starbursts and merger activity as the origin, respectively, 
synchrotron ageing of nonthermal plasma produced by AGNs can also naturally explain the steepening, 
which is proposed as a way to distinguish AGNs \citep{2011ApJ...739L..25L}.

To examine the statistical impact of the AGNs on the radio spectra through comparison with previous observations, 
and to validate the AGN diagnosis by discussing the origin of the radio spectra, 
we present the results of multifrequency observations conducted with the Karl G. Jansky Very Large Array (VLA).
While previous studies were intended for the entire set of ULIRGs \citep{1991ApJ...378...65C,2008A&A...477...95C,2010MNRAS.405..887C,2011ApJ...739L..25L,2013ApJ...777...58M}, 
in this paper, we observe ULIRGs hosting buried AGNs identified by MIR or  submillimetre spectroscopy but no evidence of AGNs at optical wavelengths \citep{2007ApJS..171...72I,2016AJ....152..218I,2018ApJ...856..143I,2019ApJS..241...19I,2014AJ....148....9I}.
We describe our sample sources in Section \ref{sec:sample}, 
the observations and data reduction in Sections \ref{sec:obs},
and the results in Section \ref{sec:res}.
Section \ref{sec:discuss} discusses the origin of the radio properties, including the spectra.
Finally, our conclusions are summarised in Section \ref{sec:conlusion}.
In this paper, we define the spectral index, $\alpha$, as $f_\nu \propto \nu^\alpha$, 
where $f_\nu$ is flux density at frequency, $\nu$.
In this definition, a steep spectrum is represented by $\alpha<0$.

% -----------------------------------------------------------------------------------
\section{Sample}\label{sec:sample}
% -----------------------------------------------------------------------------------
Table~\ref{tbl:sample} lists our sample.
We selected the sample from nearby ULIRGs that show positive AGN signatures in \citet{2007ApJS..171...72I}.
Their sample consists of 48 ULIRGs at $z<0.15$ from the IRAS 1\,Jy sample \citep{1998ApJS..119...41K}, whose optical classification is LINER or \ion{H}{2} \citep{1999ApJ...522..113V}. 
They classify the ULIRGs into four levels based on the strength of the AGN sign presented by MIR spectroscopy.
Of 26 ULIRGs with some AGN signatures, we selected nine ULIRGs whose R.A. are ranged 22 to 5\,hour.
In addition, for all but two (IRAS 00482$-$2721, IRAS 23327$+$2913) of the sample submillimetre spectroscopy is performed by \citet{2019ApJS..241...19I}, where the presence of buried AGNs is suggested by an HCN-to-HCO$^+$ ($J =$ 3--2) flux-density ratio higher than unity.
Table~\ref{tbl:AGN} shows the result of the AGN diagnostics.

In this study, we also added IRAS\,22491$-$1808, whose MIR spectral feature of the source implies a strong starburst sign, to our sample because the source is considered to contain deeply buried AGN that was elusive at even MIR \citep{2014AJ....148....9I,2016AJ....152..218I,2018ApJ...856..143I,2019ApJS..241...19I}.
IRAS\,22491$-$1808 shows a high HCN-to-HCO$^+$ flux-density ratio similar to that of AGN-dominated ULIRGs and a ratio of the vibrationally excited HCN emission line to the vibrational ground emission line higher than the starburst galaxies, which is presumed to be produced by hot dust heated by the AGN.

\begin{table*}
\begin{minipage}{150mm}
\caption{ULIRG Sample for VLA Observation.}
\label{tbl:sample}
\begin{tabular}{cccccccccccccc}
\hline															
ID  &   Object	&	R.A.	&	Dec.	&	$z$	&	$\log L_{\rm IR}$	&	Optical Class	&	$J$	&	$K$	\\
	&	&	(J2000)	&	(J2000)	&		&	($L_\odot$)	&		&	(mag)	&	(mag)	\\
(1)	&	(2)	&	(3)	&	(4)	&	(5)	&	(6)	&	(7)	&	(8)	&   (9)\\
\hline															
1	&	IRAS 00091$-$0738	&	00 11 43.25	&	$-$07 22 07.5	&	0.118	&	12.19 	&	\ion{H}{2}	&	$15.34\pm0.05$	&	$13.96\pm0.07$	\\
2	&	IRAS 00188$-$0856	&	00 21 26.48	&	$-$08 39 27.1	&	0.128	&	12.33 	&	LINER	&	$15.18\pm0.08$	&	$13.26\pm0.04$	\\
3	&	IRAS 00482$-$2721	&	00 50 40.37	&	$-$27 04 38.8	&	0.129	&	12.00 	&	LINER	&	$16.04\pm0.18$	&	$15.04\pm0.39$	\\
4	&	IRAS 01004$-$2237	&	01 02 49.92	&	$-$22 21 57.0	&	0.118	&	12.24 	&	\ion{H}{2}	&	$16.10\pm0.14$	&	$14.34\pm0.14$	\\
5	&	IRAS 01166$-$0844	&	01 19 07.54	&	$-$08 29 09.4	&	0.118	&	12.03 	&	\ion{H}{2}	&	$15.72\pm0.12$	&	$14.54\pm0.28$	\\
6	&	IRAS 01298$-$0744	&	01 32 21.41	&	$-$07 29 08.9	&	0.136	&	12.27 	&	\ion{H}{2}	&	$16.33\pm0.26$	&	$14.95\pm0.30$	\\
7	&	IRAS 04103$-$2838	&	04 12 19.47	&	$-$28 30 24.4	&	0.118	&	12.15 	&	LINER	&	$14.78\pm0.07$	&	$13.29\pm0.04$	\\
8	&	IRAS 21329$-$2346	&	21 35 45.86	&	$-$23 32 35.6	&	0.125	&	12.09 	&	LINER	&	$15.52\pm0.03$	&	$14.25\pm0.07$	\\
9	&	IRAS 22491$-$1808	&	22 51 49.23	&	$-$17 52 22.9	&	0.076	&	12.09 	&	\ion{H}{2}	&	$14.50\pm0.05$	&	$13.31\pm0.09$	\\
10	&	IRAS 23327$+$2913	&	23 35 11.92	&	$+$29 30 00.4	&	0.107	&	12.06 	&	LINER	&	$15.16\pm0.07$	&	$13.60\pm0.07$	\\
\hline															
\end{tabular}

\medskip
Columns are as follows:
(1) Object ID;
(2) IRAS source name; 
(3)(4) Optical positions; 
(5) Redshift; 
(6) Infrared luminosity; 
Reference for the columns (2)--(6) are \citet{1998ApJS..119...41K}.
(7) Optical classification \citep{1999ApJ...522..113V};
(8)(9) $J$- and $K$-band aperture magnitudes from the 2MASS all sky survey \citep{2003yCat.2246....0C,2006AJ....131.1163S}. The observed wavelengths of the $J$ and $K$ bands in the survey are 1.25 and 2.16\,$\mu$m, respectively.

\end{minipage}
\end{table*}

\begin{table*}
\begin{minipage}{120mm}
\caption{AGN diagnoses of the sample.}
\label{tbl:AGN}
\begin{tabular}{cccccccccccccc}
\hline															
Object	&	EW$_{6.2}^{\mathrm{PAH}}$	&	EW$_{7.7}^{\mathrm{PAH}}$	&	EW$_{11.3}^{\mathrm{PAH}}$	&	$\tau_{9.7}$	&	$\tau_{18}/\tau_{9.7}$	&	$\tau'_{18}/\tau_{9.7}$	&	HCN/HCO$^+$ 	\\
	&	(nm)	&	(nm)	&	(nm)	&		&		&		&	($J=$~3--2)	\\
(1)	&	(2)	&	(3)	&	(4)	&	(5)	&	(6)	&	(7)	&	(8)	\\
\hline															
IRAS 00091$-$0738	&	\phn\textbf{50}	&	350	&	255	&	\textbf{3.2}	&	0.33	&	0.39	&	\textbf{1.6~$\pm$~0.3}	\\
IRAS 00188$-$0856	&	\phn\textbf{85}	&	305	&	290	&	\textbf{2.5}	&	\textbf{0.22}	&	\textbf{0.26}	&	\textbf{1.9~$\pm$~0.2}	\\
IRAS 00482$-$2721	&	260	&	710	&	465	&	\textbf{2.1}	&	0.29	&	0.36	&	...	\\
IRAS 01004$-$2237	&	\phn\textbf{45}	&	\textbf{190}	&	\phn\phn\textbf{8}	&	0.4	&	...	&	...	&	1.0~$\pm$~0.2	\\
IRAS 01166$-$0844	&	\phn\textbf{30}	&	430	&	\textbf{190}	&	\textbf{3.0}	&	\textbf{0.25}	&	\textbf{0.30}	&	\textbf{1.6~$\pm$~0.3}	\\
IRAS 01298$-$0744	&	\phn\textbf{10}	&	380	&	\textbf{155}	&	\textbf{4.0}	&	0.27	&	0.34	&	\textbf{1.3~$\pm$~0.2}	\\
IRAS 04103$-$2838	&	\textbf{165}	&	365	&	\textbf{120}	&	0.6	&	...	&	...	&	\textbf{1.2~$\pm$~0.3}	\\
IRAS 21329$-$2346	&	230	&	635	&	370	&	\textbf{2.2}	&	...	&	...	&	0.9~$\pm$~0.2	\\
IRAS 22491$-$1808	&	325	&	695	&	460	&	1.5	&	...	&	...	&	\textbf{2.3~$\pm$~0.2}	\\
IRAS 23327$+$2913	&	255	&	485	&	\textbf{175}	&	1.3	&	...	&	...	&	1.0~$\pm$~0.2	\\
\hline															
\end{tabular}

\medskip
Columns are as follows:
(1) 
IRAS source name;
(2)--(4) 
Rest-frame equivalent width of PAH emission at 6.2, 7.7, and 11.3\,$\mu$m, respectively. 
Positive AGN signatures are indicated in bold font, based on the low equivalent width of 
EW$_{6.2}^{\mathrm{PAH}} < 180$\,nm, 
EW$_{7.7}^{\mathrm{PAH}} < 230$\,nm, and 
EW$_{11.3}^{\mathrm{PAH}} < 200$\,nm;
(5) 
Optical depth of silicate dust absorption feature at 9.7\,$\mu$m,
against a power-law continuum determined from data points at 7.1 and 14.2\,$\mu$m.
Positive AGN signatures are indicated in bold font, based on the high optical depth of $\tau_{9.7}>2$;
(6)(7)
ratio of optical depth of silicate dust absorption feature at 18\,$\mu$m to that at 9.7\,$\mu$m, which represents dust temperature gradients.
$\tau_{18}$ and $\tau'_{18}$ are the optical depths against a power-law continuum determined from data points at 14.2 and 24\,$\mu$m, and  14.2 and 29\,$\mu$m, respectively.
Positive AGN signatures are indicated in bold font, based on strong dust temperature gradients determined by $\tau_{18}/\tau_{9.7}< 0.27$ and $\tau'_{18}/\tau_{9.7} < 0.32$;
Reference for the columns (2)--(7) is \citet{2007ApJS..171...72I}.
(8)
HCN-to-HCO$^+$ ($J =$ 3--2) flux-density ratio.
Values at the nuclear peak position within the beam (typically $\sim 0.2$\,arcsec) are shown.
%and those within a 1\,arcsec diameter circular aperture (1.5\,arcsec and 2.0\,arcsec for IRAS 21329$-$2346 and IRAS 22491$-$1808, respectively) are shown.
Positive AGN signatures are indicated in bold font, based on the ratio higher than unity.
References for the column (8) are \citet{2016AJ....152..218I,2019ApJS..241...19I}.
\end{minipage}
\end{table*}

% -----------------------------------------------------------------------------------
\section{Observation and Data Reduction}\label{sec:obs}
% -----------------------------------------------------------------------------------
Our multifrequency JVLA observations were made between October 2015 and August 2016 (project codes: AH1185 and AH1196).
To make observations with a similar $u$-$v$ range at different bands, we utilised the B, C, C, and D configurations for observations at the L, C, X, and Ku bands, respectively (the D configuration is the most compact).
Table~\ref{tbl:observation} shows an observational summary.
As presented in the VLA status summary at the time, the typical largest angular scales are 60, 120, 70, and 50\,arcsec for snapshot images at the L, C, X, and Ku bands, respectively.

We reduced the data using the Common Astronomy Software Applications \citep[CASA;][]{2007ASPC..376..127M} package developed by the National Radio Astronomy Observatory (NRAO).
We calibrated the bandpass, phase, and amplitude response of individual antennas with a standard procedure.
During the calibration, we flagged bad data based on the signal-to-noise ratio of visibilities.
All data in each band were combined into a single image using CASA's multi-term multifrequency synthesis capabilities with a Briggs weighting setting {\tt robust} $=$ 0.5.
We set a reference frequencies of output images at the C, X, and Ku bands as 5.5\,GHz, 9.0\,GHz, and 14.0\,GHz, respectively, which are default values of CASA calculated as the middle of the observed frequency ranges.
For the L band, we manually set reference frequency of 1.4\,GHz to evaluate flux-density variation by comparing the flux density with prior survey observations made at the same frequency \citep{1995ApJ...450..559B,1998AJ....115.1693C}.
We also conducted phase-only self-calibration of the data whenever possible.
Table~\ref{tbl:imaging} shows the resolution and noise level of each image.

The images of each source were restored by a beam of the lowest resolution image.
We measured the flux densities of the targets by CASA's task {\tt IMFIT},
which fit an elliptical Gaussian component within a small box containing a radio source.
We estimated the flux-density errors by the root sum square of the error presented by {\tt IMFIT}, which contains thermal noise and fitting errors, and a nominal calibration uncertainty of 5\% presented in the VLA status summary at the time.

\begin{table*}
\begin{minipage}{131mm}
\caption{Observation Summary.}
\label{tbl:observation}
\begin{tabular}{cccccl}
\hline											
Band	&	Centre Frequencies	&	Bandwidth	&	Configuration	&	Date	&\multicolumn{1}{c}{Observed targets}	    \\\hline
L	&	1.25 and 1.75\,GHz	&	\phn512\,MHz each	 &	B	&	2016 July 08	&	5, 6, 7	\\
	&		&		&		&	2016 August 19	&	1, 2, 3, 4, 8, 9, 10	\\
C	&	\phn5.0 and \phn6.0\,GHz	&	1024\,MHz each	&	C	&	2016 March 19	&	1, 2, 3, 4, 5, 6, 8, 9, 10	\\
X	&	\phn8.5 and \phn9.5\,GHz	&	1024\,MHz each	&	C	&	2016 March 18	&	1, 2, 3, 4, 5, 6, 8, 9, 10	\\
Ku	&	13.0 and 15.0\,GHz	&	1024\,MHz each	&	D	&	2015 October 18	&	9, 10	\\
	&		&		&		&	2015 October 19	&	1, 2	\\
	&		&		&		&	2015 October 20	&	3, 4, 5, 6, 7, 8    \\
\hline				
\end{tabular}

\medskip
Numbers in observed targets column are object IDs in Table~\ref{tbl:sample}.
\end{minipage}
\end{table*}

\begin{table*}
\begin{minipage}{\textwidth}
\renewcommand{\tabcolsep}{0.77mm}
\caption{Synthesised beam and r.m.s noise level of the images.}
\label{tbl:imaging}
\begin{tabular}{ccccccccccccccccccc}
\hline		
Object	&	\multicolumn{3}{c}{1.4\,GHz}					&&	\multicolumn{3}{c}{5.5\,GHz}					&&	\multicolumn{3}{c}{9.0\,GHz}					&&	\multicolumn{3}{c}{14.0\,GHz}					\\\cline{2-4}\cline{6-8}\cline{10-12}\cline{14-16}
	&	Beam size	&	P.A.	&	r.m.s	&&	Beam size	&	P.A.	&	r.m.s	&&	Beam size	&	P.A.	&	r.m.s	&&	Beam size	&	P.A.	&	r.m.s	\\
	&	(arcsec$^2$)	&	(deg)	&	($\mu$Jy\,beam$^{-1}$)	&&	(arcsec$^2$)	&	(deg)	&	($\mu$Jy\,beam$^{-1}$)	&&	(arcsec$^2$)	&	(deg)	&	($\mu$Jy\,beam$^{-1}$)	&&	(arcsec$^2$)	&	(deg)	&	($\mu$Jy\,beam$^{-1}$)	\\\hline
IRAS 00091$-$0738	&	\phn7.2~$\times$~3.9	&	140	&	\phn67	&&	4.8~$\times$~3.3	&	174	&	25 	&&	3.2~$\times$~2.2	&	163	&	19 	&&	\phn9.9~$\times$~4.5	&	134	&	18 	\\
IRAS 00188$-$0856	&	\phn7.8~$\times$~3.9	&	141	&	\phn78	&&	5.0~$\times$~3.5	&	173	&	16 	&&	3.5~$\times$~2.0	&	149	&	16 	&&	\phn9.6~$\times$~4.4	&	135	&	17 	\\
IRAS 00482$-$2721	&	13.6~$\times$~3.6	&	148	&	\phn49	&&	8.3~$\times$~3.2	&	167	&	16 	&&	5.5~$\times$~2.0	&	158	&	16 	&&	10.1~$\times$~4.5	&	\phn\phn4	&	13 	\\
IRAS 01004$-$2237	&	13.0~$\times$~4.7	&	143	&	\phn71	&&	7.9~$\times$~3.6	&	167	&	18 	&&	4.9~$\times$~2.1	&	157	&	17 	&&	10.2~$\times$~4.8	&	\phn\phn4	&	17 	\\
IRAS 01166$-$0844	&	\phn5.8~$\times$~4.6	&	\phn31	&	\phn51	&&	5.4~$\times$~3.7	&	153	&	15 	&&	3.6~$\times$~2.1	&	151	&	14 	&&	\phn7.1~$\times$~5.3	&	\phn15	&	12 	\\
IRAS 01298$-$0744	&	\phn5.7~$\times$~3.6	&	\phn41	&	\phn82	&&	5.3~$\times$~3.4	&	157	&	15 	&&	3.7~$\times$~2.1	&	146	&	14 	&&	\phn6.9~$\times$~5.0	&	\phn19	&	12 	\\
IRAS 04103$-$2838	&	\phn8.1~$\times$~4.0	&	\phn\phn1	&	\phn62	&&	$\dots$	&	$\dots$	&	$\dots$	&&	$\dots$	&	$\dots$	&	$\dots$	&&	12.1~$\times$~4.7	&	162	&	15 	\\
IRAS 21329$-$2346	&	\phn7.9~$\times$~4.7	&	164	&	199 	&&	8.0~$\times$~3.4	&	\phn21	&	17 	&&	4.4~$\times$~2.1	&	\phn10	&	16 	&&	\phn8.4~$\times$~3.9	&	169	&	17 	\\
IRAS 22491$-$1808	&	\phn7.3~$\times$~4.1	&	144	&	\phn74	&&	6.2~$\times$~3.8	&	\phn\phn9	&	17 	&&	3.7~$\times$~2.1	&	174	&	15 	&&	\phn7.8~$\times$~4.2	&	154	&	18 	\\
IRAS 23327$+$2913	&	\phn4.9~$\times$~3.7	&	125	&	\phn67	&&	4.4~$\times$~3.4	&	\phn36	&	19 	&&	2.2~$\times$~2.1	&	\phn45	&	15 	&&	\phn6.2~$\times$~5.5	&	110	&	17 	\\
\hline
\end{tabular}

\end{minipage}
\end{table*}

% -----------------------------------------------------------------------------------
\section{Results}\label{sec:res}
% -----------------------------------------------------------------------------------
\subsection{Radio Spectrum and Morphology}\label{sec:spectrum}
\begin{figure*}
	\centering
	\includegraphics[width=0.42\linewidth]{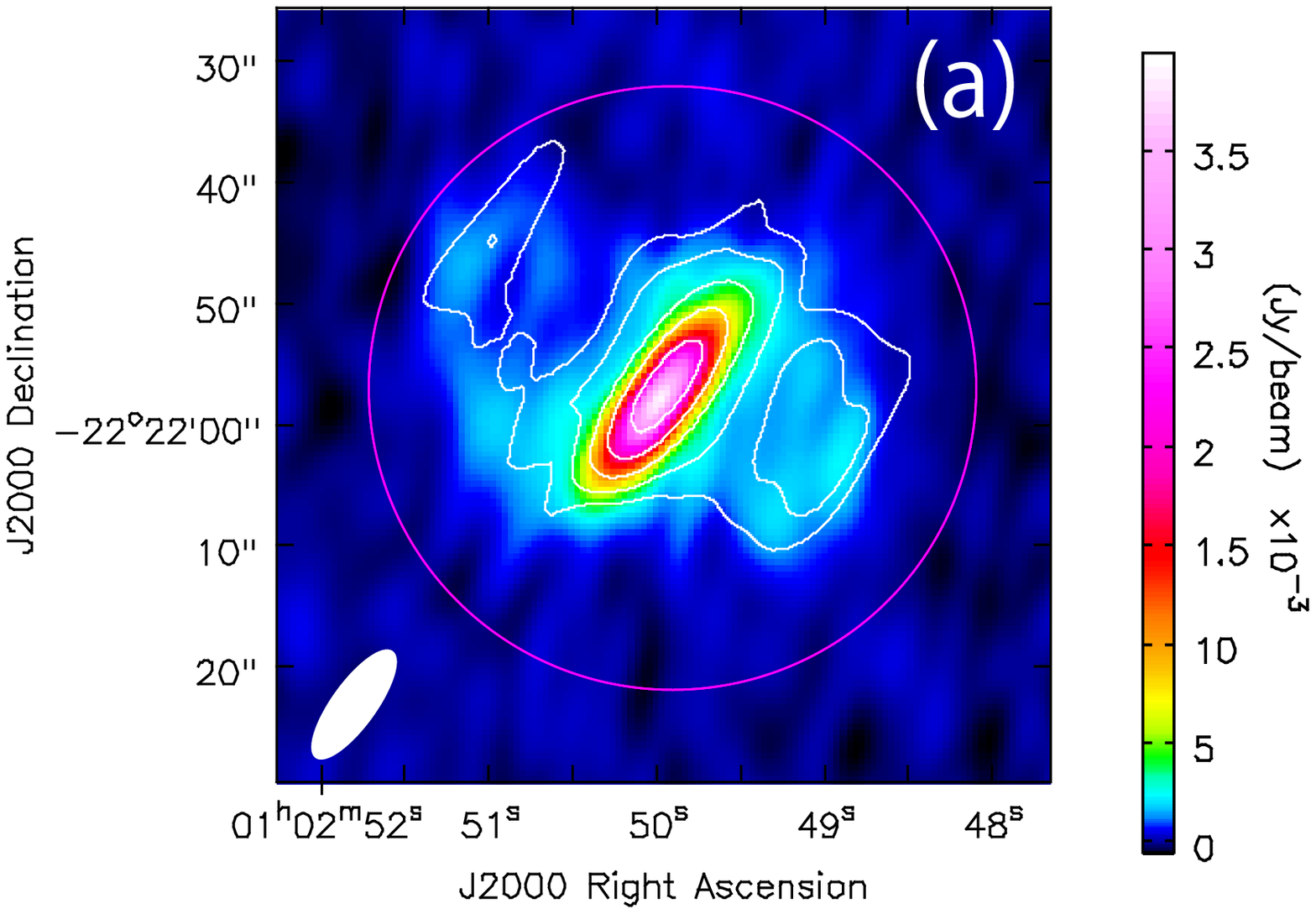}\hspace{1em}
	\includegraphics[width=0.42\linewidth]{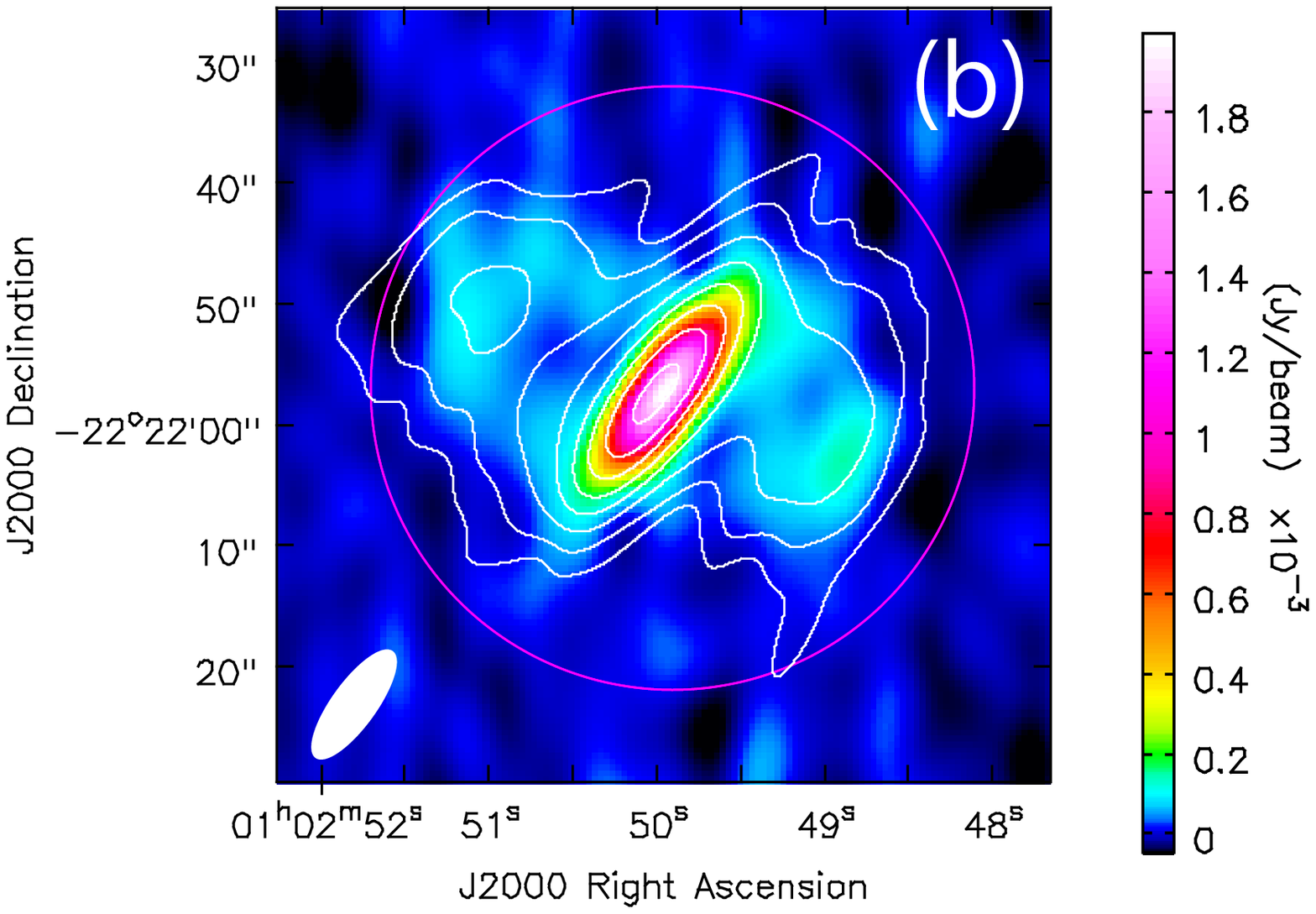}
	\caption{
	Radio images of IRAS\,01004$-$2237 convolved with a synthesised beam of the observation at 1.4\,GHz that is $13.0 \times 4.7$\,arcsec$^2$ at a position angle of 143\,degree. 
	Circles coloured by magenta with a radius of 50\,arcsec are regions for obtaining integrated flux density of extended emission.
	(a) Image at 5.5\,GHz with 1.4-GHz contours overlaid.
	Contour levels begin at 231\,$\mu$Jy\,beam$^{-1}$ (3$\sigma$ noise level of the image) and are spaced by a factor of 2. 
	(b) Image at 14.0\,GHz with 9.0-GHz contours overlaid.
	Contour levels begin at 58\,$\mu$Jy\,beam$^{-1}$ (3$\sigma$ noise level of the image) and are spaced by a factor of 2.
	}
	\label{fig:im-IRAS01004}
\end{figure*}

% -----------------------------------------------------------------------------------
Table~\ref{tbl:result} lists the results of the flux-density measurements and derived spectral indices of our targets.
In images at all bands, all targets but one show a pointlike structure.
The one exception is IRAS\,01004$-$2237, whose image is displayed in Figure~\ref{fig:im-IRAS01004}.
The radio emission of the source spreads symmetrically over $\sim$\,50\,arcsec toward the southwest and northeast.
Because 1\,arcsec corresponds to $\sim 2$\,kpc at the redshift, 
its projected linear size is $\sim 100$\,kpc.
The values of IRAS\,01004$-$2237 shown in Table~\ref{tbl:result} are the result of the core.
Also note that, since IRAS\,04103$-$2838 was observed only in the L and Ku bands, we only obtained a spectral index between 1.4 and 14.0\,GHz, whose value is $-0.87\pm 0.03$.

\citet{2013ApJ...768....2M} have reported that ULIRGs with extremely low 6.2-$\mu$m PAH equivalent width of $\mathrm{EW}_{6.2}^\mathrm{PAH} < 100\,\mathrm{nm}$, which suggests the presence of buried AGNs, did not have a spectral index between 1.5 and 8.4\,GHz steeper than $-0.6$.
By contrast, we find that IRAS\,00188$-$0856, whose $\mathrm{EW}_{6.2}^\mathrm{PAH}$ is 85\,nm, shows both $\alpha^{1.4}_{5.5}$ and $\alpha^{5.5}_{9.0}$ steeper than $-0.8$

% figure: map of 01004 (extended)

% table: flux densities and SPIX
\begin{table*}
\begin{minipage}{\textwidth}
\renewcommand{\tabcolsep}{1.4mm}
\caption{Flux densities and spectral indices of cores obtained by our new VLA observations.}
\label{tbl:result}
\begin{tabular}{cccccccccccccc}
\hline																					
Object	&		Convolved beam		&	P.A.	&	$f_{1.4}$	&	$f_{5.5}$	&	$f_{9.0}$	&	$f_{14.0}$	&	$\alpha^{1.4}_{5.5}$	&	$\alpha^{5.5}_{9.0}$	&	$\alpha^{9.0}_{14.0}$	\\
	&		(arcsec$^2$)		&	(deg)	&	(mJy)	&	(mJy)	&	(mJy)	&	(mJy)	&		&		&		\\
(1)	&		(2)		&	(3)	&	(4)	&	(5)	&	(6)	&	(7)	&	(8)	&	(9)	&	(10)	\\
\hline																				
IRAS 00091$-$0738	&	\phn9.9~$\times$~4.5	&	134	&	\phn$4.73\pm0.34$	&	\phn$3.60\pm0.19$	&	\phn$2.96\pm0.15$	&	\phn$2.03\pm0.11$	&	$-0.20\pm0.07$	&	$-0.40\pm0.15$	&	$-0.85\pm0.17$	\\
IRAS 00188$-$0856	&	\phn9.6~$\times$~4.4	&	135	&	$20.26\pm1.23$	&	\phn$6.70\pm0.34$	&	\phn$4.32\pm0.22$	&	\phn$2.74\pm0.14$	&	$-0.81\pm0.06$	&	$-0.89\pm0.14$	&	$-1.03\pm0.16$	\\
IRAS 00482$-$2721	&	13.6~$\times$~3.6	&	149	&	\phn$6.54\pm0.35$	&	\phn$2.67\pm0.14$	&	\phn$1.75\pm0.09$	&	\phn$1.28\pm0.07$	&	$-0.65\pm0.05$	&	$-0.86\pm0.15$	&	$-0.71\pm0.17$	\\
IRAS 01004$-$2237	&	13.0~$\times$~4.7	&	143	&	\phn$8.40\pm0.50$	&	\phn$4.79\pm0.31$	&	\phn$3.25\pm0.19$	&	\phn$2.45\pm0.15$	&	$-0.41\pm0.06$	&	$-0.79\pm0.18$	&	$-0.64\pm0.19$	\\
IRAS 01166$-$0844	&	\phn7.1~$\times$~5.3	&	\phn15	&	\phn$1.27\pm0.13$	&	\phn$0.72\pm0.04$	&	\phn$0.42\pm0.02$	&	\phn$0.43\pm0.03$	&	$-0.41\pm0.09$	&	$-1.07\pm0.16$	&	$+0.03\pm0.21$	\\
IRAS 01298$-$0744	&	\phn6.9~$\times$~5.0	&	\phn19	&	\phn$4.22\pm0.53$	&	\phn$3.51\pm0.18$	&	\phn$3.15\pm0.16$	&	\phn$2.61\pm0.13$	&	$-0.14\pm0.10$	&	$-0.22\pm0.14$	&	$-0.43\pm0.16$	\\
IRAS 04103$-$2838	&	12.1~$\times$~4.7	&	162	&	$13.35\pm0.75$	&	$\dots$	&	$\dots$	&	\phn$1.79\pm0.09$	&	$\dots$	&	$\dots$	&	$\dots$	\\
IRAS 21329$-$2346	&	\phn8.4~$\times$~3.9	&	169	&	\phn$9.34\pm0.70$	&	\phn$3.34\pm0.17$	&	\phn$2.14\pm0.11$	&	\phn$1.59\pm0.08$	&	$-0.75\pm0.07$	&	$-0.90\pm0.15$	&	$-0.67\pm0.17$	\\
IRAS 22491$-$1808	&	\phn7.8~$\times$~4.2	&	154	&	\phn$6.62\pm0.40$	&	\phn$4.05\pm0.21$	&	\phn$3.28\pm0.17$	&	\phn$2.66\pm0.14$	&	$-0.36\pm0.06$	&	$-0.43\pm0.15$	&	$-0.47\pm0.16$	\\
IRAS 23327$+$2913	&	\phn6.2~$\times$~5.5	&	110	&	\phn$8.46\pm0.46$	&	\phn$3.26\pm0.16$	&	\phn$2.12\pm0.11$	&	\phn$1.49\pm0.08$	&	$-0.70\pm0.05$	&	$-0.87\pm0.14$	&	$-0.81\pm0.16$	\\\hline
average	&		$\dots$		&	$\dots$	&	$\dots$	&	$\dots$	&	$\dots$	&	$\dots$	&	$-0.54\pm0.02$	&	$-0.71\pm0.05$	&	$-0.65\pm0.06$	\\
\hline																				
\end{tabular}

\medskip
Columns are as follows:
(1) IRAS source name; 
(2)(3) Convolved beam size and position angle; 
(4)--(7) Flux densities at 1.4, 5.5, 9.0, and 14.0\,GHz, respectively;
(8)--(10) Spectral indices between neighbouring bands.
We obtained flux densities only at 1.4 and 14.0\,GHz for IRAS\,04103$-$2838, whose spectral index between the frequencies is $-0.87\pm 0.03$.
The average of spectral indices weighted by the measurement error is also presented in each frequency range, whose error represents standard error on the mean.
\end{minipage}
\end{table*}

\begin{figure*}
	\centering
	\includegraphics[width=0.32\linewidth]{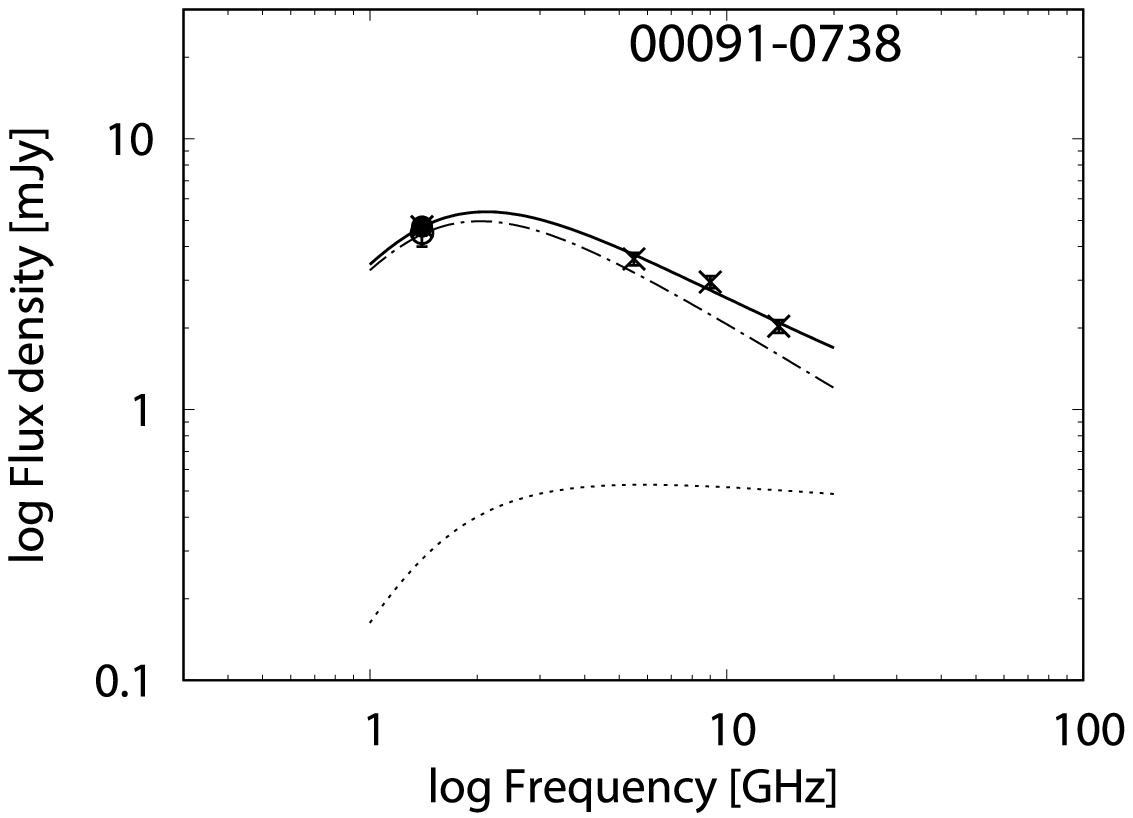}
	\includegraphics[width=0.32\linewidth]{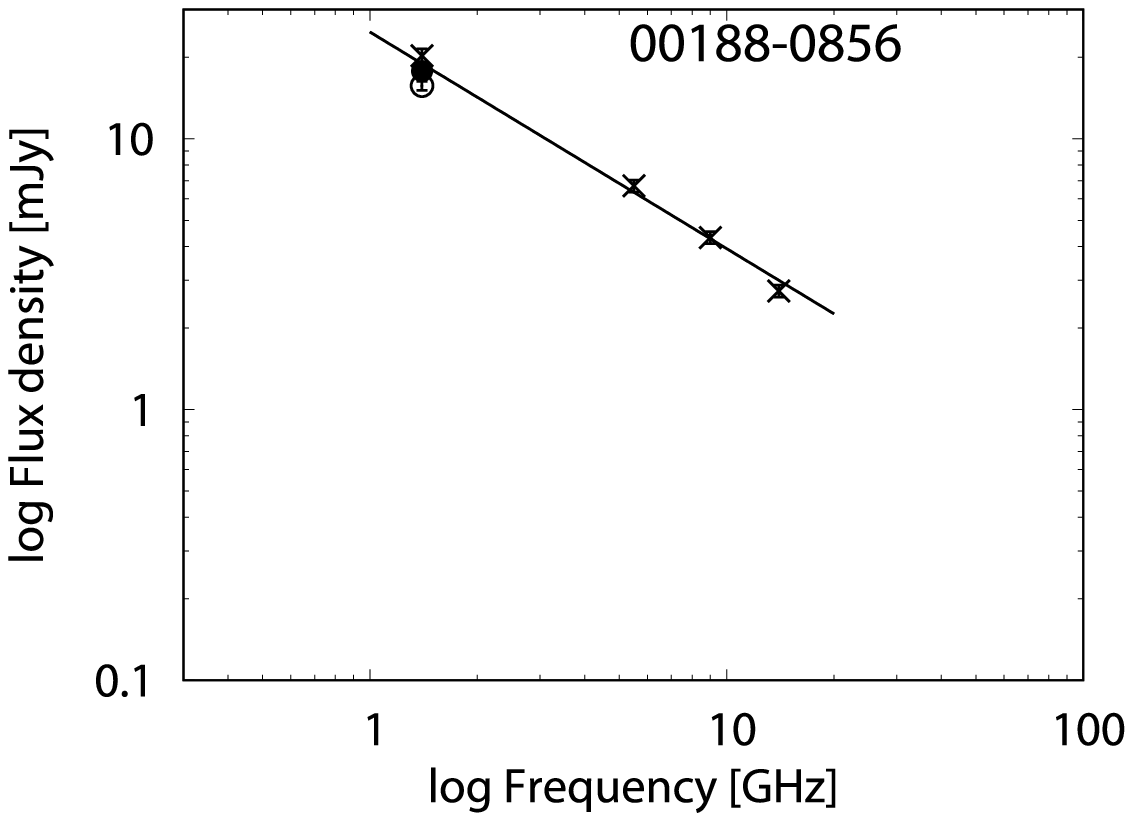}
	\includegraphics[width=0.32\linewidth]{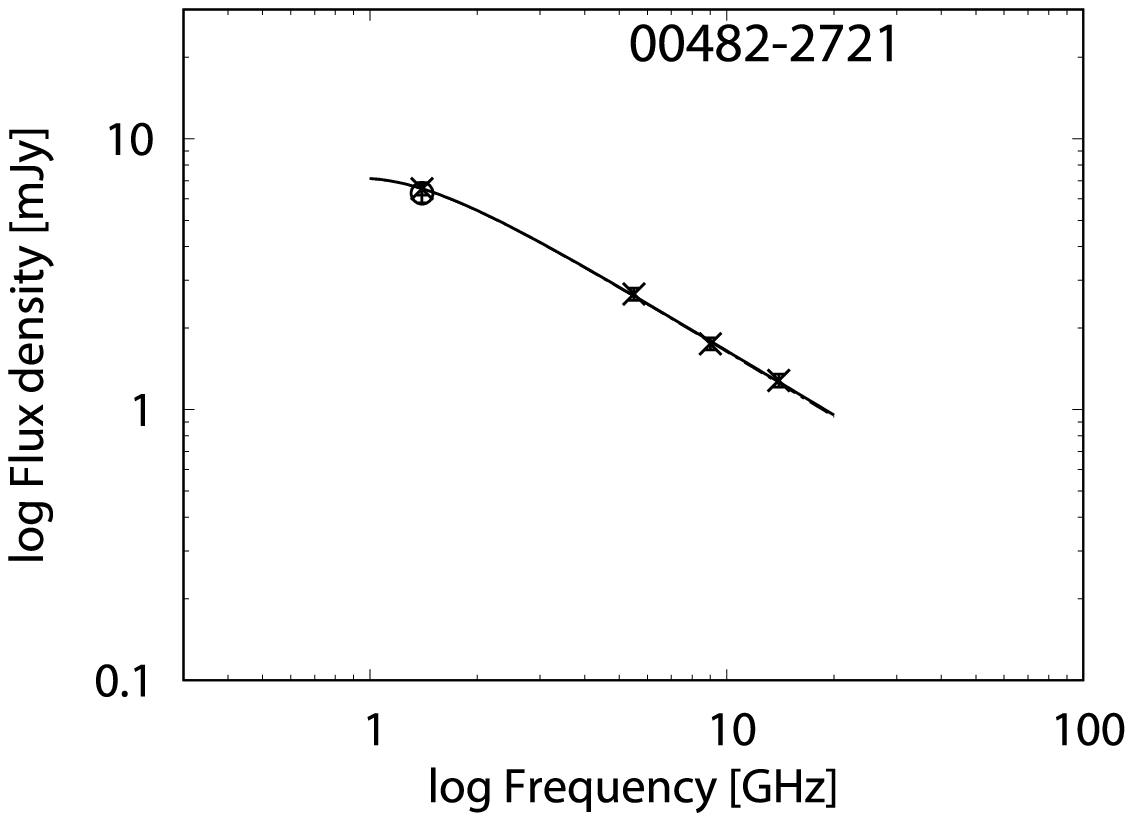}\\[0.5em]
	\includegraphics[width=0.32\linewidth]{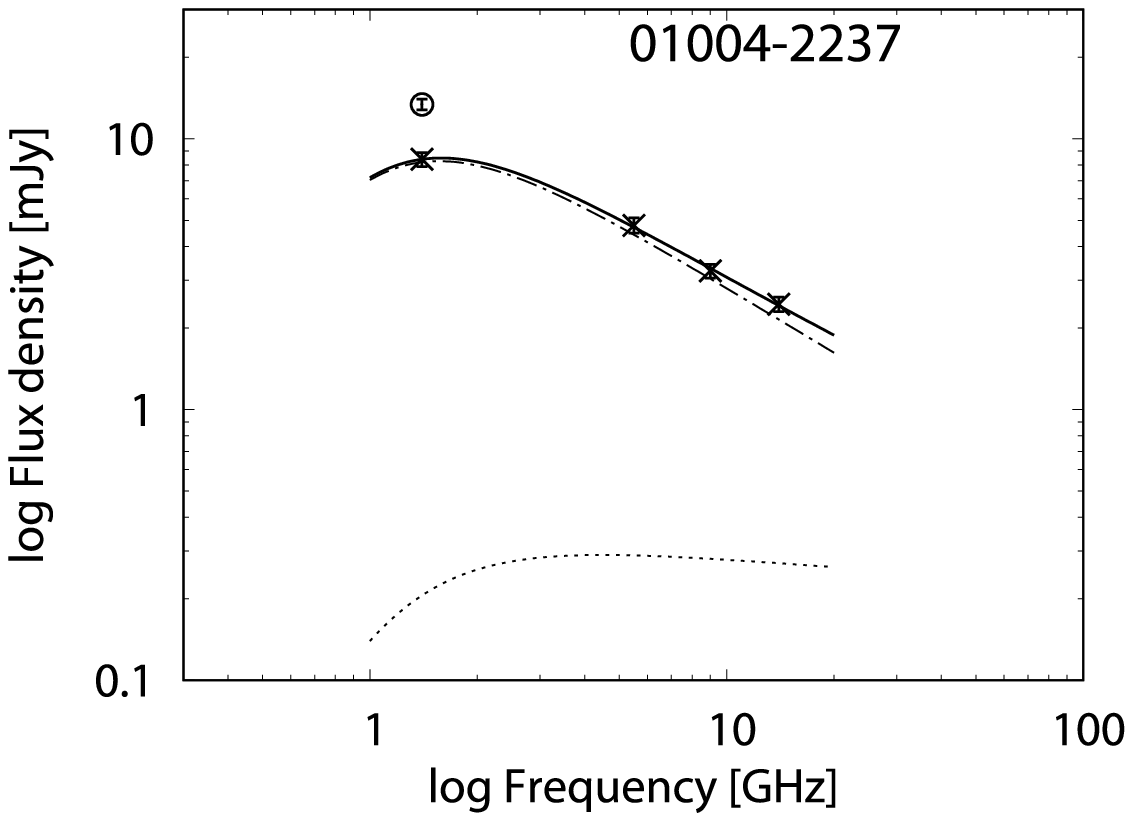}
	\includegraphics[width=0.32\linewidth]{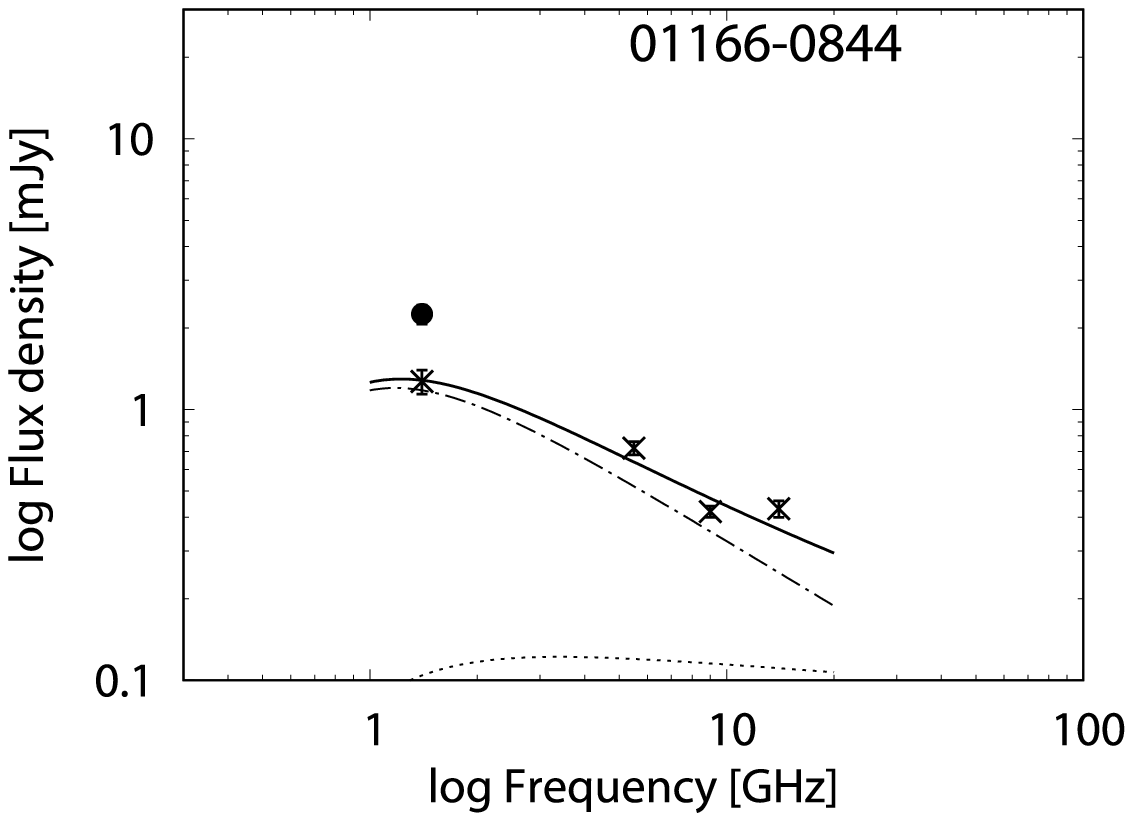}
	\includegraphics[width=0.32\linewidth]{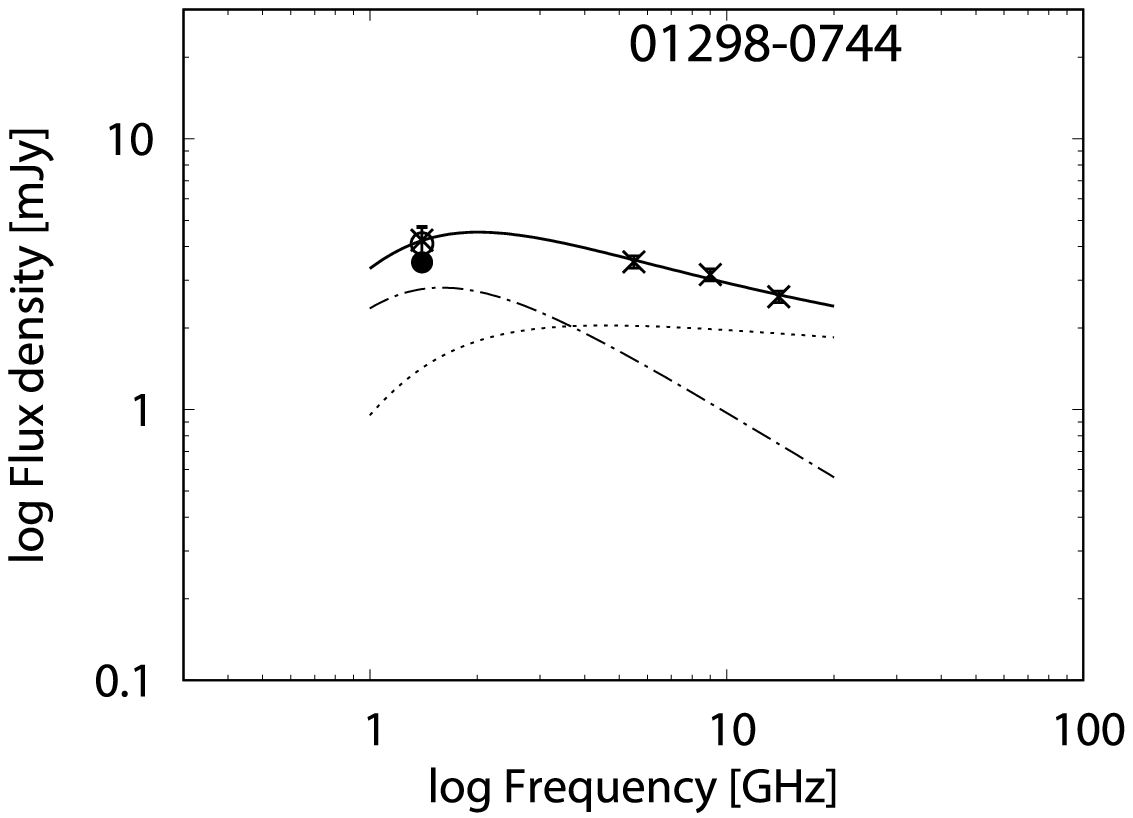}\\[0.5em]
	\includegraphics[width=0.32\linewidth]{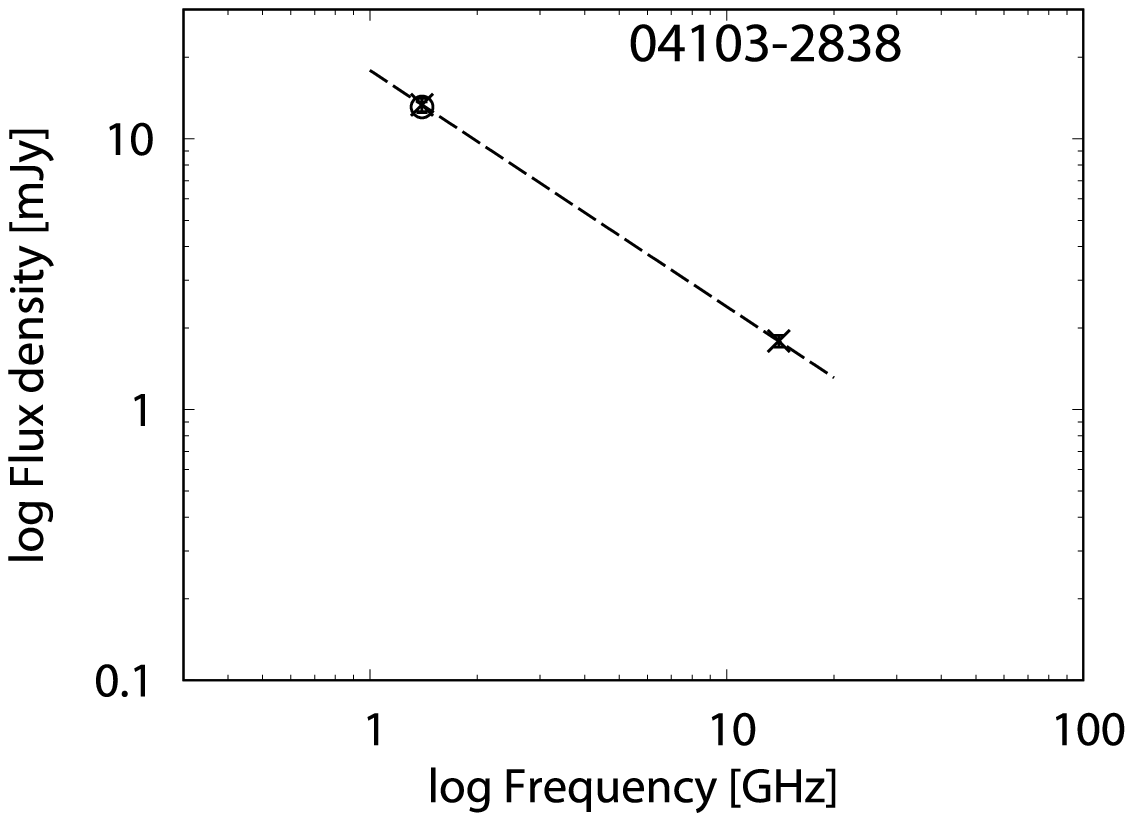}
	\includegraphics[width=0.32\linewidth]{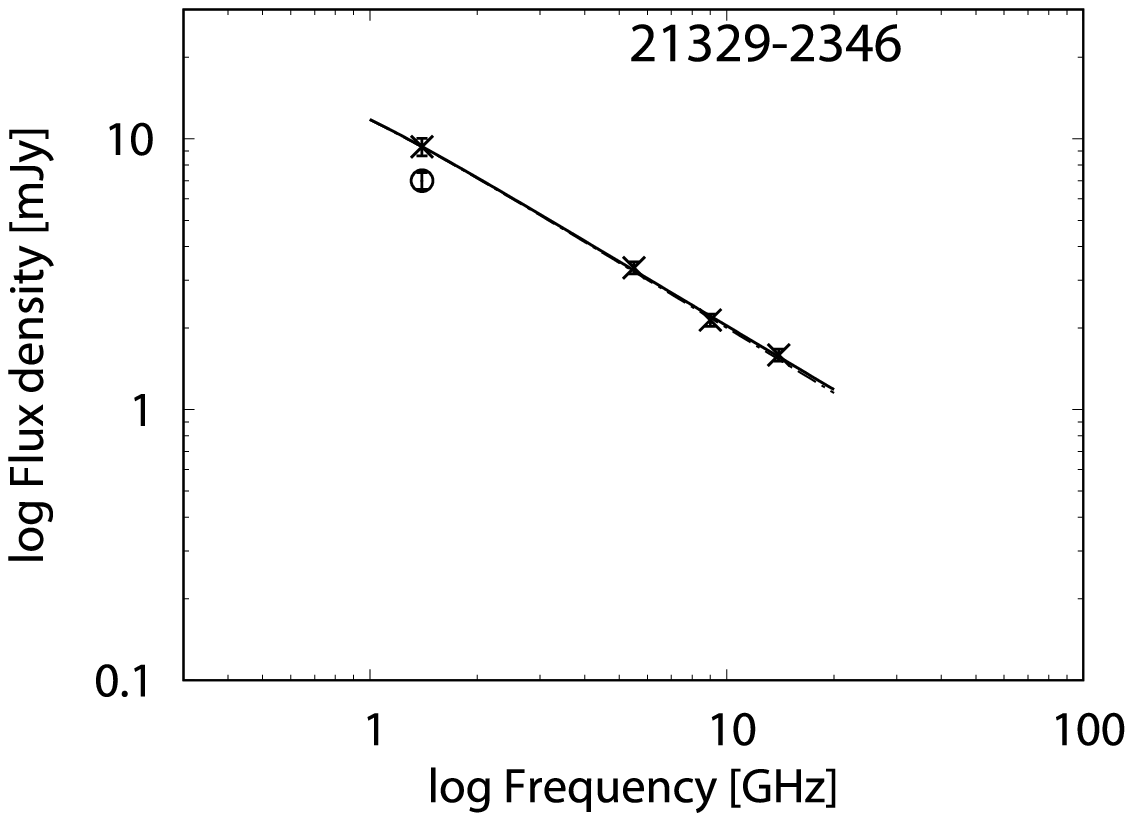}
	\includegraphics[width=0.32\linewidth]{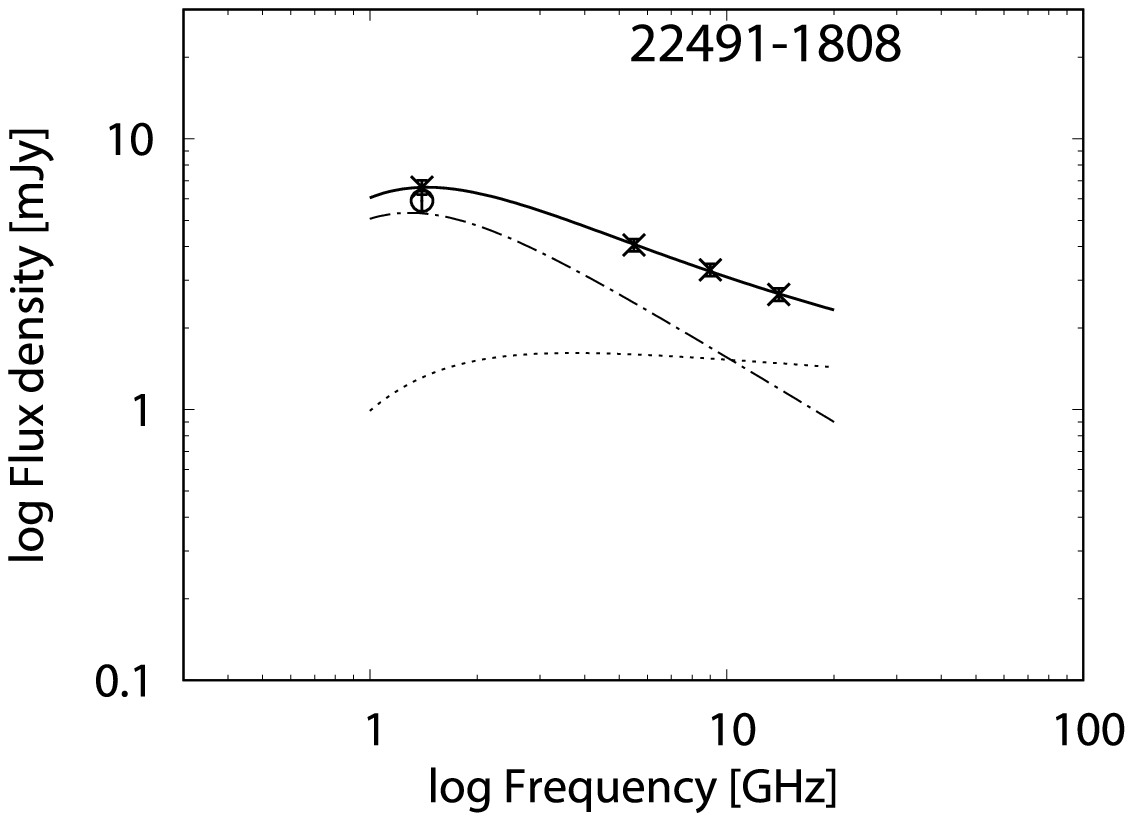}\\[0.5em]
	\includegraphics[width=0.32\linewidth]{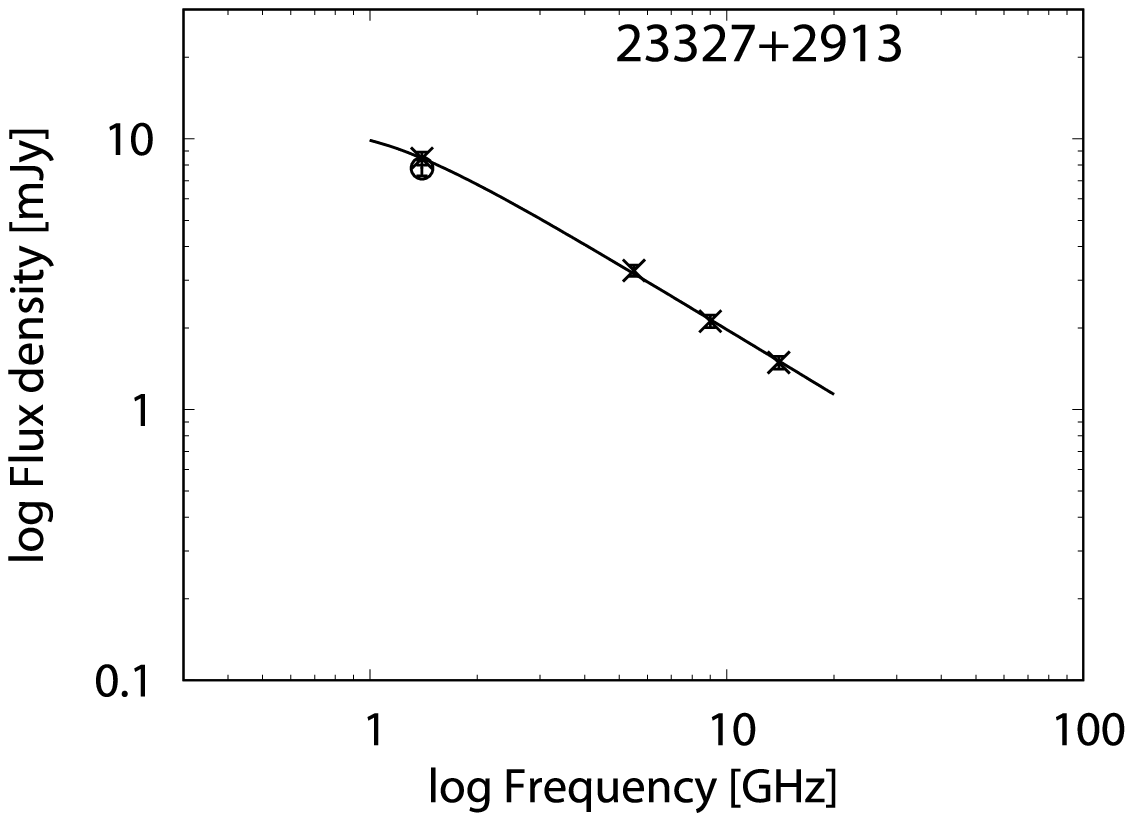}
	\caption{
	    Radio spectra of the targets obtained by our observations.
		 Observed data (crosses) and best-fit model for the spectra presented by  Equation~\ref{eq:model} (solid line) are indicated.
		We also show data points from the FIRST survey (filled circles; \citealt{1995ApJ...450..559B}) and NVSS (open circles; \citealt{1998AJ....115.1693C}).
		 The model has a nonthermal component absorbed by thermal plasma with free-free emission.
		The thermal component (dotted line) and the nonthermal component (dot-dashed line) are also indicated.
		For IRAS\,04103$-$2838, we made observations only at two bands, which is insufficient to fit the model. 
		Hence, a simple power law fit is indicated (dashed line).
		Because IRAS\,00188$-$0856, IRAS\,00482$-$2721, IRAS\,21329$-$2346, and IRAS\,23327$+$2913 show a small thermal fraction, their thermal components are not illustrated in the panels, and the solid line overlaps the dot-dashed line.
	}
	\label{fig:spectra}
\end{figure*}

Figure~\ref{fig:spectra} displays radio spectra of the targets, where data points obtained by previous survey observations (NVSS; \citealt{1998AJ....115.1693C}, and the FIRST survey; \citealt{1995ApJ...450..559B}) are also indicated.
Generally, radio emission of starburst galaxies comes from a mixture of thermal and nonthermal plasma suffering from absorption by thermal plasma.
Thus, assuming a uniform free-free optical depth, the radio brightness of a starburst galaxy is modelled by
\begin{eqnarray}
f_\nu \propto T_\mathrm{e}(1-e^{-\tau_\nu}) \Bigl[ 1+\frac{1}{H}\Bigl(\frac{\nu}{\mathrm{GHz}}\Bigl)^{0.1+\alpha_0} \Bigl]\nu^2,
\label{eq:model}
\end{eqnarray}
where $T_\mathrm{e}$ and $\tau_\nu\propto \nu^{-2.1}$ are the electron temperature and optical depth of the thermal plasma, respectively, and $H$ is a ratio of the thermal to nonthermal components \citep{1991ApJ...378...65C}.
In this model, if $\tau_\nu \ll 1$, then we obtained $(1-e^{-\tau_\nu}) \sim \tau_\nu$.
Thus, $f_\nu \sim \nu^{-0.1}$ if the thermal component is dominant, while $f_\nu \sim \nu^{\alpha}$ if the thermal component is negligible.
We fit this model to the obtained spectra, which is also shown in Figure~\ref{fig:spectra}.
Because our observations provide only four data points, we fix $\alpha_0 = -0.8$, which is a typical value of nonthermal synchrotron emission.
In the figure, we also illustrate the thermal and nonthermal components of the model.

\begin{figure}
	\centering
	\includegraphics[width=0.7\linewidth]{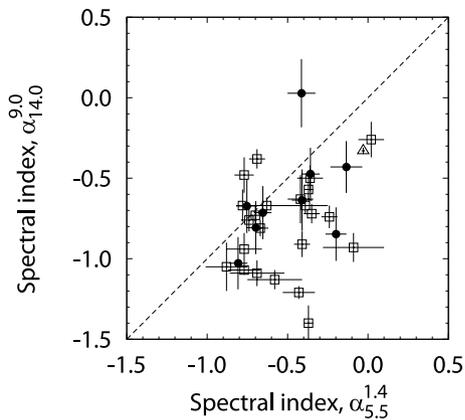}\\
	\caption{
    	A comparison between low- and high-frequency spectral indices.
    	The dashed line indicates where the spectral index is same at both frequency ranges.
    	Filled circles indicate our results, which are based on spectral indices between 1.4 and 5.5\,GHz, and between 9.0 and 14.0\,GHz ($\alpha^{1.4}_{5.5}$ and $\alpha^{9.0}_{14.0}$, respectively).
    	The Pearson correlation coefficient is 0.447 with a p-value of 0.228.
    	Results of the entire ULIRGs \citep{2008A&A...477...95C} are also shown by open squares, which are based on spectral indices between 1.4 and 4.8\,GHz, and between 8.4 and 22.5\,GHz.
    	An open triangle shows Mrk\,231 provided by the same author.
    	Excluding Mrk\,231, the Pearson correlation coefficient is 0.272 with a p-value of 0.178.
    }
	\label{fig:SPIX-SPIX}
\end{figure}

We also provide the average of spectral indices in each frequency range in Table~\ref{tbl:result}.
The spectral index between two frequencies, $\nu_1$ and $\nu_2$, is now described as $\alpha^{\nu_1}_{\nu_2}$.
We find that the average $\alpha^{1.4}_{5.5}$ is significantly less negative (i.e. flatter) than $\alpha^{5.5}_{9.0}$, while no significant difference is found between $\alpha^{5.5}_{9.0}$ and $\alpha^{9.0}_{14.0}$.
That is, flattening of the spectrum is discovered only in the lowest frequency range but not in the higher frequency range.
Figure~\ref{fig:SPIX-SPIX} shows a comparison between $\alpha^{1.4}_{5.5}$ and $\alpha^{9.0}_{14.0}$, which suggests no significant correlation at 95\% significance. 
These trends are similar to previous reports for the entire ULIRGs \citep{2008A&A...477...95C,2013ApJ...777...58M,2018MNRAS.474..779G}, which will be discussed in Section~\ref{sec:comparison}.

\begin{figure}
	\centering
	\includegraphics[width=0.75\linewidth]{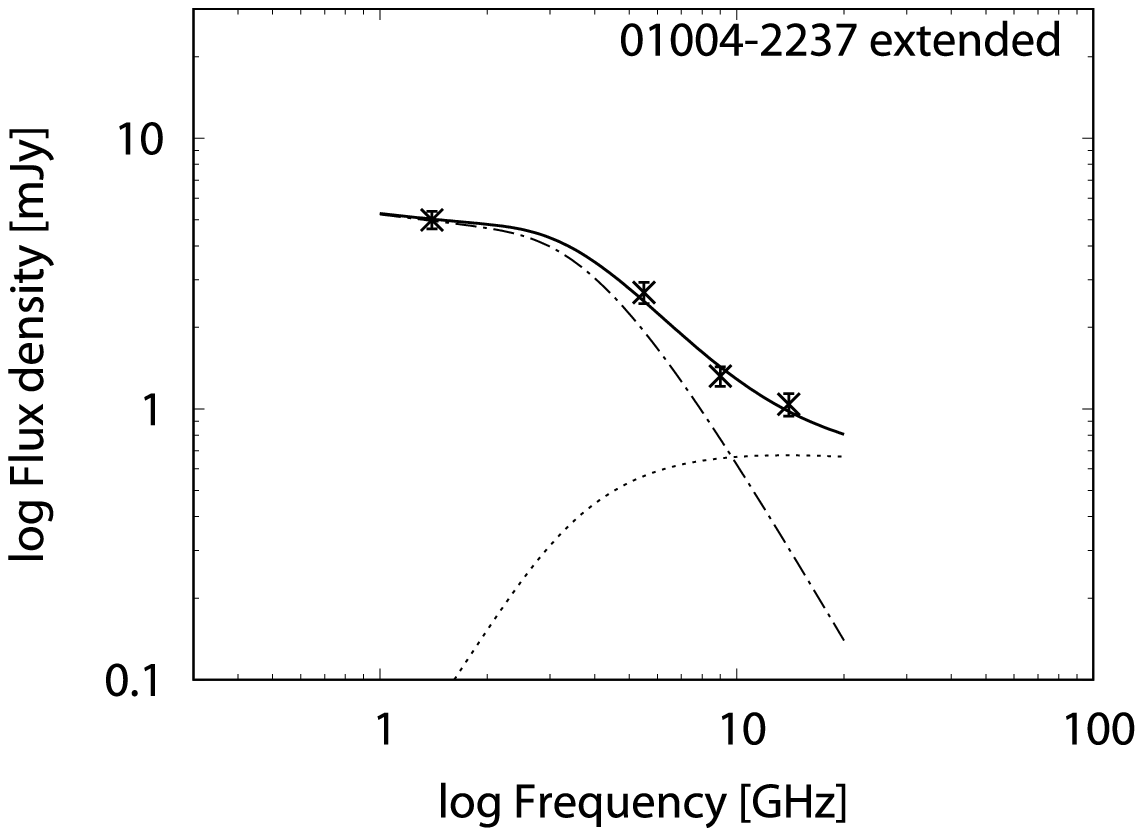}
	\caption{
	    Radio spectrum of extended emission from IRAS\,01004$-$2237 obtained by our observations.
		 Observed data (crosses) and best-fit model for the spectrum presented by  Equation~\ref{eq:model} (solid line) are indicated.
		 Model has a nonthermal component absorbed by thermal plasma with free-free emission.
		Thermal component (dotted line) and nonthermal component (dot-dashed line) are also indicated.
		}
	\label{fig:spectra_IRAS01004_extended}
\end{figure}

In Figure~\ref{fig:spectra_IRAS01004_extended}, we show the radio spectrum of the extended emission associated with IRAS\,01004$-$2237.
Flux-density measurements were made by subtracting the core emission obtained by {\tt IMFIT} from the integrated flux density within $50$\,arcsec of the centre (see also Figure~\ref{fig:im-IRAS01004}).
We also illustrate the best-fit model provided by Equation~\ref{eq:model}.
We will discuss the origin of this feature in Section~\ref{sec:01004}

% figure: spectra

% figure: aLC-aXU

% figure: spectrum of 01004 (extended)

% -----------------------------------------------------------------------------------
\subsection{Flux-density variation}\label{sec:variability}
% -----------------------------------------------------------------------------------
To evaluate the significance of flux-density variability, we utilise the following index introduced by \citet{2006ApJ...639..716Z},
\begin{eqnarray}
\sigma_\mathrm{var} = \frac{|S_1-S_2|}{\sqrt{\sigma_1^2+\sigma_2^2}},
\end{eqnarray}
where $S_i$ and $\sigma_i$ are the flux density and uncertainty of the $i$-th epoch data, respectively.
Objects with $\sigma_\mathrm{var} > 3$ are candidates for variable sources \citep{2006ApJ...639..716Z,2007ApJ...661L.139G}.
We compare the flux density at 1.4\,GHz obtained by our new observations with those from previous survey observations (NVSS and the FIRST survey).
Both survey observations do not present the flux-density errors with calibration uncertainty, which now we assume 5\% of the total flux density as we did for our observations.
When we compare data measured with different array configurations, sources that suggest peak flux density at a high resolution larger than integrated flux density at lower resolution are considered as possible variable objects.
Now, because our observations and the FIRST survey utilised the B configuration while NVSS was conducted with the D configuration, the resolution of our observation ($\sim$5\,arcsec) is comparable with the FIRST survey, but higher than NVSS ($\sim$45\,arcsec).

Table~\ref{tbl:variability} presents the results.
While IRAS\,01004$-$2237 showing the extended emission (Figure\,\ref{fig:im-IRAS01004}) indicates $\sigma_\mathrm{var} > 3$ in comparison with NVSS, this large deviation comes from the effect of array configurations.
On the other hand, the other comparisons show $\sigma_\mathrm{var} < 3$.
Although these two-epoch comparisons are not enough to conclude that the sources are stable, we do not find significant flux-density variation for our targets.
This finding suggests that, while our multiband observations were not made simultaneously, radio spectra and spectral indices presented in Section \ref{sec:spectrum} are physically meaningful for the following discussions.

% table: variability
\begin{table*}
\begin{minipage}{120mm}
\caption{Significance of flux-density variability.}
\label{tbl:variability}
\begin{tabular}{cccccccccc}
\hline											
Object	&	$f_{1.4}^{\rm NVSS}$	&	$I_{1.4}^{\rm FIRST}$	&	NVSS-FIRST	&	ours--NVSS	&	ours--FIRST	\\
	&	(mJy)	&	(mJy\,beam$^{-1}$)	&		&		&		\\
(1)	&	(2)	&	(3)	&	(4)	&	(5)	&	(6)	\\
\hline											
IRAS 00091$-$0738	&	\phn$4.5\pm0.9$	&	\phn$5.0\pm0.4$	&	0.78	&	0.34	&	0.34	\\
IRAS 00188$-$0856	&	$15.7\pm1.2$	&	$17.2\pm1.6$	&	0.84	&	1.97	&	1.60	\\
IRAS 00482$-$2721	&	\phn$6.3\pm0.9$	&	$\dots$	&	$\dots$	&	0.05	&	$\dots$	\\
IRAS 01004$-$2237	&	$13.4\pm1.1$	&	$\dots$	&	$\dots$	&	6.77	&	$\dots$	\\
IRAS 01166$-$0844	&	$\dots$	&	\phn$1.3\pm0.3$	&	$\dots$	&	$\dots$	&	2.82	\\
IRAS 01298$-$0744	&	\phn$4.1\pm0.9$	&	\phn$3.9\pm0.8$	&	0.15	&	1.21	&	0.91	\\
IRAS 04103$-$2838	&	$13.1\pm1.1$	&	$\dots$	&	$\dots$	&	0.51	&	$\dots$	\\
IRAS 21329$-$2346	&	\phn$7.0\pm0.9$	&	$\dots$	&	$\dots$	&	0.25	&	$\dots$	\\
IRAS 22491$-$1808	&	\phn$5.9\pm0.9$	&	$\dots$	&	$\dots$	&	0.16	&	$\dots$	\\
IRAS 23327$+$2913	&	\phn$7.8\pm0.9$	&	$\dots$	&	$\dots$	&	0.55	&	$\dots$	\\
\hline											
\end{tabular}

\medskip
Columns are as follows:
(1) IRAS source name; 
(2) Integrated flux density at 1.4\,GHz provided by NVSS \citep{1998AJ....115.1693C}; 
(3) Peak flux density at 1.4\,GHz provided by the FIRST survey \citep{1995ApJ...450..559B}; 
(4)--(6) Significance of variability between NVSS and the FIRST survey, our observations and NVSS, and our observations and the FIRST survey, respectively.
\end{minipage}
\end{table*}

% -----------------------------------------------------------------------------------
\subsection{Radio spectral indices and FIR-to-radio ratio}\label{sec:q}
% -----------------------------------------------------------------------------------
Table~\ref{tbl:q} shows FIR-to-radio ratio, $q_\nu$, and related FIR properties obtained by IRAS observations.
%\citep{1989AJ.....98..766S}
Here, $q_\nu$ is defined as follows:
$q_\nu = \log[(f_{\mathrm{FIR}}/3.75 \times 10^{12}\,\mathrm{Hz})/f_\nu]$, 
where $f_{\mathrm{FIR}}$ is the FIR flux density in units of W\,m$^{-2}$\,Hz$^{-1}$ commonly defined as $f_\mathrm{FIR}=1.26\times10^{14} (2.58 \times f_{60}+f_{100})$ \citep{1985ApJ...298L...7H}.
To compare FIR emission with intrinsic radio power, we utilise flux density at 9.0\,GHz, where emission is free from absorption in most of the sources (see Figure~\ref{fig:spectra}).
Table~\ref{tbl:q} also lists FIR-to-radio ratio at 1.4\,GHz after correction for FFA, $q_{1.4}'$, introduced by \citet{1991ApJ...378...65C}, which is now defined as $q_{1.4}' = q_{1.4}+(\alpha_0-\alpha^{1.4}_{5.5})\log (5.5/1.4)$.
They find that galaxies with large infrared luminosities of $\gtrsim 10^{12}L_{\odot}$ have a distribution of $q_{1.4}'$ narrower than that of $q_{1.4}$.
Our sample shows a similar trend, where the scatter of $q_{1.4}$ and $q_{1.4}'$ are 0.30 and 0.22, respectively.
Now, the scatter of $q_{9.0}$ is 0.23, which suggests that $q_{9.0}$ behaves in the same way as $q_{1.4}'$. 
Since $q_{9.0}$ is calculated without assuming the true value of $\alpha_0$, we will use $q_{9.0}$ in our discussion.

\begin{figure}
	\centering
	\includegraphics[width=0.7\linewidth]{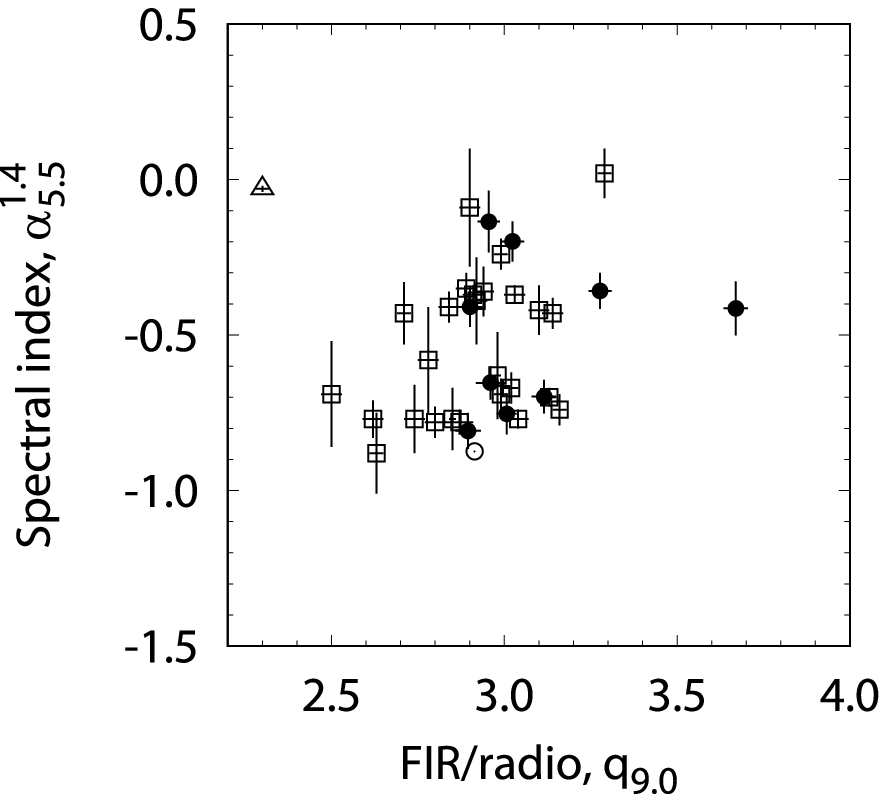}\\
	\includegraphics[width=0.7\linewidth]{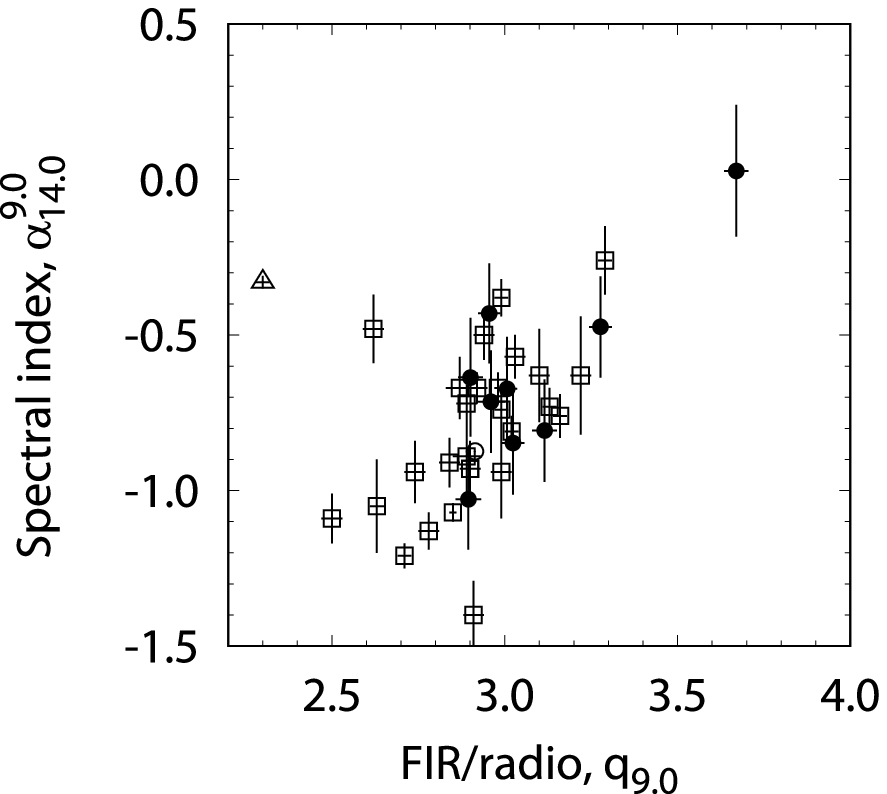}
	\caption{
    	A comparison between FIR-to-radio ratio and radio spectral indices.
    	Filled circles indicate our results, which are based on FIR-to-radio ratio at 9.0\,GHz, $q_{9.0}$.
    	Data points indicated by open circles represent IRAS\,04103$-$2838, 
    	whose spectral indices and $q_{9.0}$ were calculated by interpolation from flux densities at 1.4 and 14.0\,GHz.
    	We exclude this source to calculate following statistical values.
    	%, and spectral indices between 1.4 and 5.5\,GHz, and between 9.0 and 14.0\,GHz ($\alpha^{1.4}_{5.5}$ and $\alpha^{9.0}_{14.0}$, respectively).
    	Results of the entire ULIRGs \citep{2008A&A...477...95C} are also shown by open squares, which are based on FIR-to-radio ratio at 8.4\,GHz.
    	An open triangle shows Mrk\,231 provided by the same author, which is excluded to calculate following statistical values.
    	%, which are based on spectral indices between 1.4 and 4.8\,GHz, and between 8.4 and 22.4\,GHz.
    	\textbf{(Top)}
    	A comparison with low-frequency data.
    	Our results are based on spectral index between 1.4 and 5.5\,GHz, $\alpha^{1.4}_{5.5}$, while previous ones are based on spectral index between 1.4 and 4.8\,GHz.
    	Pearson correlation coefficients for our and previous data are 0.179 and 0.408 with p-values of 0.038 and 0.645, respectively.
    	The difference between these coefficients is not significant with p-value of 0.581.
    	\textbf{(Bottom)}
    	A comparison with high-frequency data.
    	Our results are based on spectral index between 9.0 and 14.0\,GHz, $\alpha^{9.0}_{14.0}$, while previous ones are based on spectral index between 8.4 and 22.5\,GHz.
    	Pearson correlation coefficients for our and previous data are 0.806 and 0.518 with p-values of 0.005 and 0.009, respectively.
    	The difference between these coefficients is not significant with p-value of 0.234.
    }
	\label{fig:q-SPIX}
\end{figure}

Figure~\ref{fig:q-SPIX} shows the plot of the radio spectral indices against $q_{9.0}$.
We find a significant correlation between $\alpha^{9.0}_{14.0}$ and $q_{9.0}$ at the 95\% confidence level, while no significant correlation is found if $\alpha^{1.4}_{5.5}$ is considered instead.
Thus, a large $q_{9.0}$ is associated with sources showing $\alpha^{9.0}_{14.0}\sim 0$, which suggests a flat spectrum.
In the figure, we also show data points for the entire ULIRGs  provided by \citet{2008A&A...477...95C}.
Although their observed frequencies are different from ours, both results suggest that correlation with FIR-to-radio ratio is significantly stronger at higher frequencies than at lower frequencies.

% table: FIR and NIR properties and q
\begin{table*}
\begin{minipage}{171mm}
\caption{FIR-to-radio ratios.}
\label{tbl:q}
\begin{tabular}{cccccccccccc}
\hline															
Object	&	$f_{12}$	&	$f_{25}$	&	$f_{60}$	&	$f_{100}$	&	FIR	&	$q_{1.4}$	&	$q_{9.0}$	&	$q_{1.4}'$	\\
	&	(Jy)	&	(Jy)	&	(Jy)	&	(Jy)	&	($/10^{-13}$\,W\,m$^{-2}$)	&		&		&		\\
	&	(1)	&	(2)	&	(3)	&	(4)	&	(5)	&	(6)	&	(7)	&	(8)	\\
\hline																	
IRAS 00091$-$0738	&	$0.07\pm0.02$	&	$0.22\pm0.05$	&	$2.63\pm0.18$	&	$2.52\pm0.20$	&	$1.17\pm0.07$	&	$2.82\pm0.04$	&	$3.02\pm0.03$	&	$2.46\pm0.07$	\\
IRAS 00188$-$0856	&	$0.12\pm0.04$	&	$0.37\pm0.07$	&	$2.59\pm0.23$	&	$3.40\pm0.34$	&	$1.27\pm0.09$	&	$2.22\pm0.04$	&	$2.89\pm0.04$	&	$2.23\pm0.06$	\\
IRAS 00482$-$2721	&	$0.11\pm0.03$	&	$0.18\pm0.05$	&	$1.13\pm0.14$	&	$1.84\pm0.17$	&	$0.60\pm0.05$	&	$2.39\pm0.04$	&	$2.96\pm0.04$	&	$2.30\pm0.06$	\\
IRAS 01004$-$2237	&	$0.23\pm0.07$	&	$0.66\pm0.08$	&	$2.29\pm0.16$	&	$1.79\pm0.18$	&	$0.97\pm0.06$	&	$2.49\pm0.04$	&	$2.90\pm0.04$	&	$2.26\pm0.06$	\\
IRAS 01166$-$0844	&	$0.14\pm0.04$	&	$0.18\pm0.04$	&	$1.74\pm0.10$	&	$1.42\pm0.24$	&	$0.74\pm0.05$	&	$3.20\pm0.05$	&	$3.67\pm0.04$	&	$2.97\pm0.09$	\\
IRAS 01298$-$0744	&	$0.12\pm0.03$	&	$0.28\pm0.08$	&	$2.47\pm0.15$	&	$2.08\pm0.25$	&	$1.07\pm0.06$	&	$2.83\pm0.06$	&	$2.95\pm0.03$	&	$2.43\pm0.10$	\\
IRAS 04103$-$2838	&	$0.08\pm0.02$	&	$0.54\pm0.04$	&	$1.82\pm0.09$	&	$1.71\pm0.17$	&	$0.81\pm0.04$	&	$2.21\pm0.03$	&	$2.91\pm0.03$	&	$2.25\pm0.04$	\\
IRAS 21329$-$2346	&	$0.08\pm0.02$	&	$0.16\pm0.04$	&	$1.65\pm0.08$	&	$2.22\pm0.18$	&	$0.82\pm0.03$	&	$2.37\pm0.04$	&	$3.01\pm0.03$	&	$2.34\pm0.07$	\\
IRAS 22491$-$1808	&	$0.12\pm0.04$	&	$0.55\pm0.07$	&	$5.44\pm0.38$	&	$4.45\pm0.36$	&	$2.33\pm0.13$	&	$2.97\pm0.04$	&	$3.28\pm0.03$	&	$2.71\pm0.06$	\\
IRAS 23327$+$2913	&	$0.06\pm0.02$	&	$0.23\pm0.06$	&	$2.10\pm0.13$	&	$2.81\pm0.45$	&	$1.04\pm0.07$	&	$2.51\pm0.04$	&	$3.11\pm0.04$	&	$2.45\pm0.06$	\\\hline
average	&	$\dots$	&	$\dots$	&	$\dots$	&	$\dots$	&	$\dots$	&	$2.64\pm0.01$	&	$3.09\pm0.01$	&	$2.46\pm0.02$	\\
%mean	&	$\dots$	&	$\dots$	&	$\dots$	&	$\dots$	&	$\dots$	&	$2.64\pm0.30$	&	$3.09\pm0.23$	&	$2.46\pm0.22$	\\
\hline															
\end{tabular}

\medskip
Columns are as follows:
(1) IRAS source name; 
(2)--(5) Flux densities at 12, 25, 60, and 100\,$\mu$m, respectively \citep{1990IRASF.C......0M};
(6) FIR flux density as used to calculate the FIR-to-radio ratio, $q$;
(7)(8) FIR-to-radio ratios, $q$, at 1.4, and 9.0\,GHz, respectively;
(9) FIR-to-radio ratio at 1.4\,GHz after correction for FFA, $q_{1.4}'$, introduced by \citet{1991ApJ...378...65C}. 
The average of FIR-to-radio ratios weighted by the measurement error is also presented, whose error represents standard error on the mean.
FIR-to-radio ratio of IRAS\,04103$-$2838 in each column is calculated by interpolation from flux densities at 1.4 and 14.0\,GHz, as required, but we exclude the values when the average are calculated.
\end{minipage}
\end{table*}

% figure: q and SPIX

% -----------------------------------------------------------------------------------
\subsection{Radio spectral indices and FIR colour}\label{sec:FIR}
\begin{figure}
	\centering
	\includegraphics[width=0.7\linewidth]{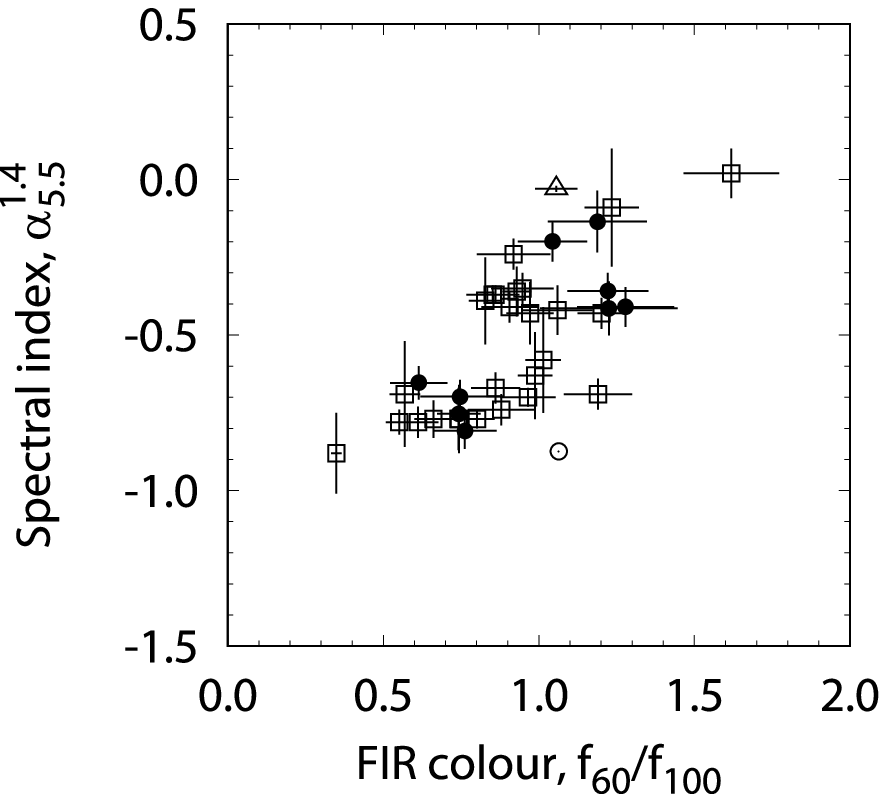}\\
	\includegraphics[width=0.7\linewidth]{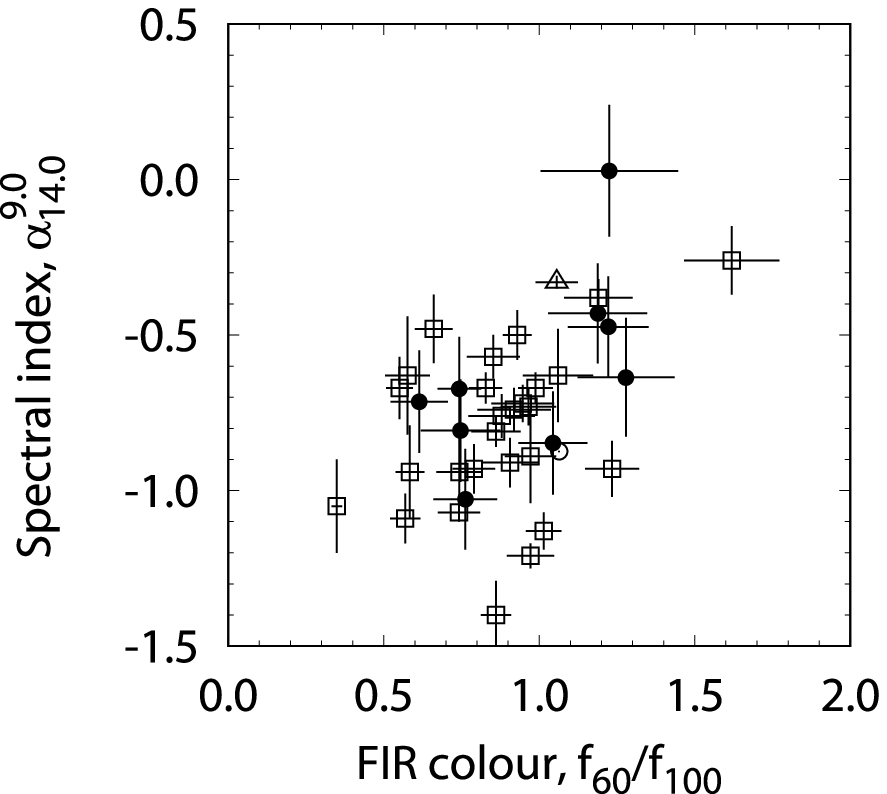}
	\caption{
    	A comparison between FIR colour, $f_{60}/f_{100}$, and radio spectral indices.
    	Filled circles indicate our results.
    	Data points indicated by open circles represent IRAS\,04103$-$2838, 
    	whose spectral indices were calculated by interpolation from flux densities at 1.4 and 14.0\,GHz.
    	We exclude this source to calculate following statistical values.
    	%, and spectral indices between 1.4 and 5.5\,GHz, and between 9.0 and 14.0\,GHz ($\alpha^{1.4}_{5.5}$ and $\alpha^{9.0}_{14.0}$, respectively).
    	Results of the entire ULIRGs \citep{2008A&A...477...95C} are also shown by open squares.
    	An open triangle shows Mrk\,231 provided by the same author, which is excluded to calculate following statistical values.
    	%, which are based on spectral indices between 1.4 and 4.8\,GHz, and between 8.4 and 22.4\,GHz.
    	\textbf{(Top)}
    	A comparison with low-frequency data.
    	Our results are based on spectral index between 1.4 and 5.5\,GHz, $\alpha^{1.4}_{5.5}$, while previous ones are based on spectral index between 1.4 and 4.8\,GHz.
    	Pearson correlation coefficients for our and previous data are 0.793 and 0.730 with p-values of 0.011 and $<$0.001, respectively.
    	The difference between these coefficients is not significant with p-value of 0.740.
    	\textbf{(Bottom)}
    	A comparison with high-frequency data.
    	Our results are based on spectral index between 9.0 and 14.0\,GHz, $\alpha^{9.0}_{14.0}$, while previous ones are based on spectral index between 8.4 and 22.5\,GHz.
    	Pearson correlation coefficients for our and previous data are 0.618 and 0.412 with p-values of 0.076 and 0.029, respectively.
    	The difference between these coefficients is not significant with p-value of 0.532.
	  }
	\label{fig:FIR-SPIX}
\end{figure}

% -----------------------------------------------------------------------------------
Figure~\ref{fig:FIR-SPIX} shows the plot of the radio spectral indices against the FIR colour, $f_{60}/f_{100}$, from IRAS observations.
We find a significant correlation between $\alpha^{1.4}_{5.5}$ and $f_{60}/f_{100}$ at the 95\% confidence level, while no correlation is found if $\alpha^{9.0}_{14.0}$ is considered instead.
Thus, a large $f_{60}/f_{100}$ is associated with  sources showing $\alpha^{1.4}_{5.5}\sim 0$, which suggests a flat spectrum.
Because $f_{60}/f_{100}$ is a measure of dust temperature heated by starbursts, 
whose large value indicates high dust temperature and then active star formation, ULIRGs with strong starbursts are accompanied by a flat spectrum at lower frequencies more often than at higher frequencies.
In the figure, we also show data points for the entire ULIRGs provided by \citet{2008A&A...477...95C}.
Although their observed frequencies are different from ours, both results suggest that correlation with FIR colour is significantly stronger at lower frequencies than at higher frequencies.

% figure: alpha and FIR colour

% -----------------------------------------------------------------------------------
\subsection{Radio spectral indices and near-infrared colour}\label{sec:NIR}
\begin{figure}
	\centering
	\includegraphics[width=0.7\linewidth]{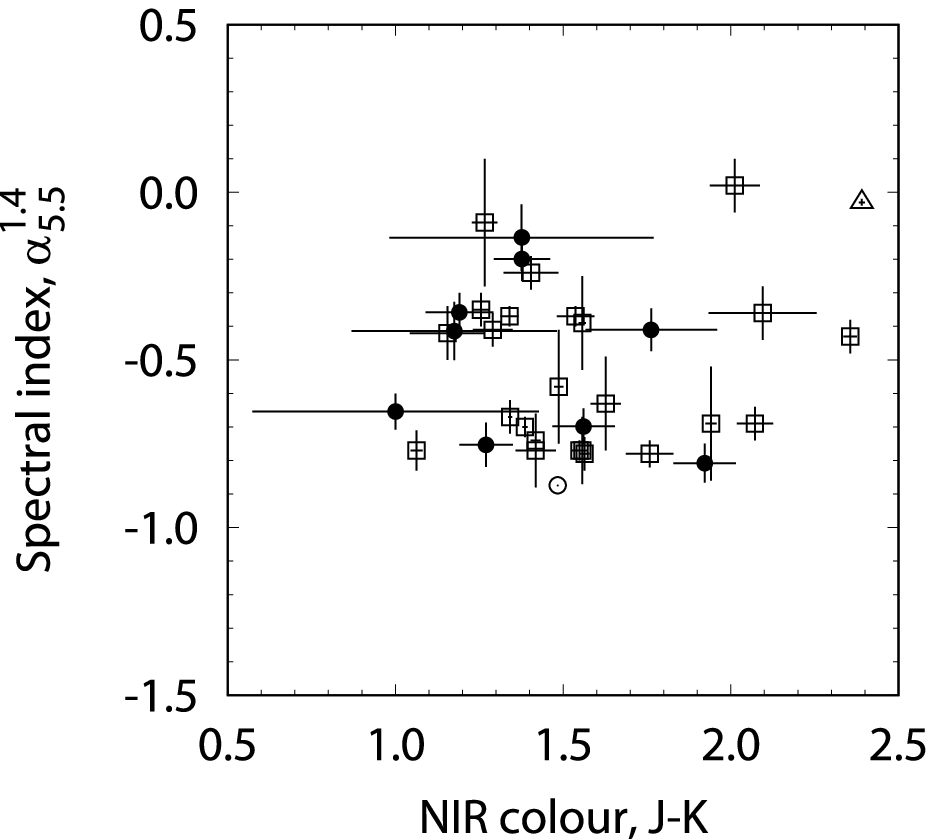}\\
	\includegraphics[width=0.7\linewidth]{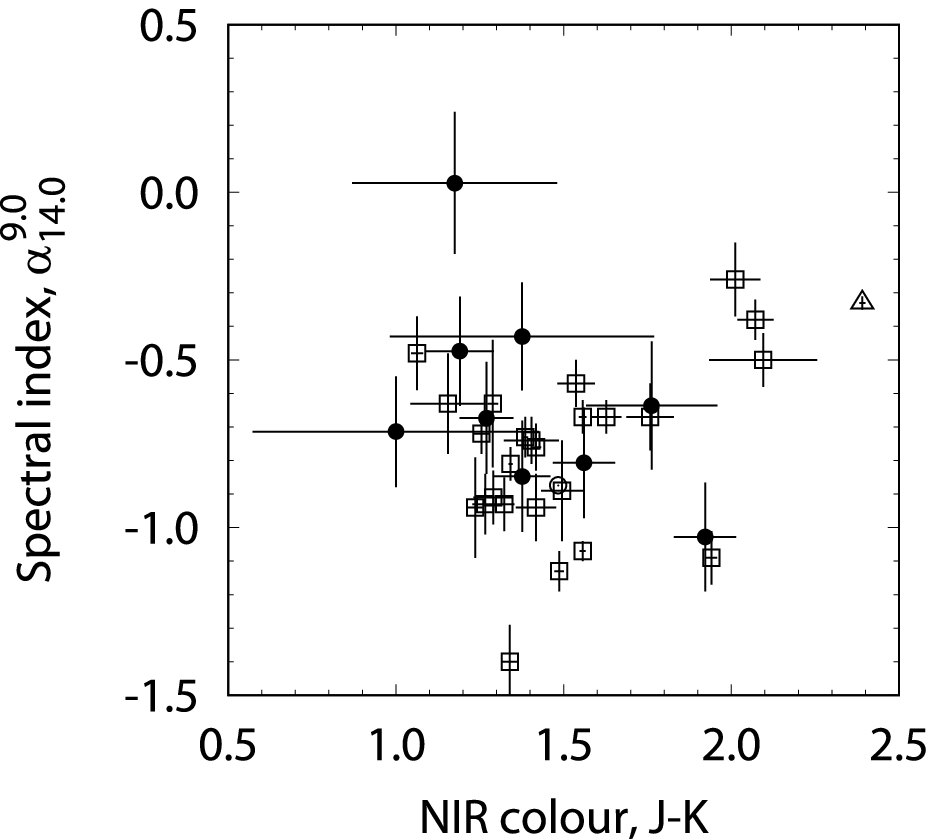}
	\caption{
    	A comparison between NIR colour, $(J-K)$, and radio spectral indices.
    	Filled circles indicate our results.
    	Data points indicated by open circles represent IRAS\,04103$-$2838, 
    	whose spectral indices were calculated by interpolation from flux densities at 1.4 and 14.0\,GHz.
    	We exclude this source to calculate following statistical values.
    	%, and spectral indices between 1.4 and 5.5\,GHz, and between 9.0 and 14.0\,GHz ($\alpha^{1.4}_{5.5}$ and $\alpha^{9.0}_{14.0}$, respectively).
    	Results of the entire ULIRGs \citep{2008A&A...477...95C} are also shown by open squares.
    	An open triangle shows Mrk\,231 provided by the same author, which is excluded to calculate following statistical values.
    	%, which are based on spectral indices between 1.4 and 4.8\,GHz, and between 8.4 and 22.4\,GHz.
    	\textbf{(Top)}
    	A comparison with low-frequency data.
    	Our results are based on spectral index between 1.4 and 5.5\,GHz, $\alpha^{1.4}_{5.5}$, while previous ones are based on spectral index between 1.4 and 4.8\,GHz.
    	Pearson correlation coefficients for our and previous data are $-0.225$ and $-0.290$ with p-values of 0.561 and 0.151, respectively.
    	The difference between these coefficients is not significant with p-value of 0.879.
    	\textbf{(Bottom)}
    	A comparison with high-frequency data.
    	Our results are based on spectral index between 9.0 and 14.0\,GHz, $\alpha^{9.0}_{14.0}$, while previous ones are based on spectral index between 8.4 and 22.5\,GHz.
    	Pearson correlation coefficients for our and previous data are $-0.526$ and $-0.080$ with p-values of 0.145 and 0.686, respectively.
	}
	\label{fig:NIR-SPIX}
\end{figure}

% -----------------------------------------------------------------------------------
In Figure~\ref{fig:NIR-SPIX}, we plot the radio spectral indices against the near-infrared (NIR) colour index, $(J-K)$, derived from the results of the 2MASS survey \citep{2006AJ....131.1163S}.
At the 95\% confidence level, the radio spectral indices in both low and high frequencies, $\alpha^{1.4}_{5.5}$ and $\alpha^{9.0}_{14.0}$, are not significantly correlated with $(J-K)$ that becomes large when galaxies shroud hot dust of 150-250\,K heated by AGNs.
In Figure~\ref{fig:FIR-SPIX}, we also show data points for the entire ULIRGs provided by \citet{2008A&A...477...95C}.
Although their observed frequencies are different from ours, both results suggest that correlation with NIR colour is not significant at both high and low frequencies.

% figure: alpha and NIR colour

% -----------------------------------------------------------------------------------
\section{Discussions}\label{sec:discuss}
% -----------------------------------------------------------------------------------
\subsection{Comparison with overall ULIRGs sample}\label{sec:comparison}
% -----------------------------------------------------------------------------------
Our new VLA observations provide spectral properties of ULIRGs whose presence of AGNs is suggested by MIR or submillimetre spectroscopy but no sign of AGN at optical wavelengths \citep{2007ApJS..171...72I,2016AJ....152..218I,2018ApJ...856..143I,2019ApJS..241...19I,2014AJ....148....9I}.
Previously, \citet{2008A&A...477...95C} made multifrequency observations for the overall ULIRG sample selected from the IRAS Bright Galaxy Sample \citep{1989AJ.....98..766S}.
A comparison between their results and ours would provide new insight to reveal the origin of the radio emission and the heating sources of the dust in ULIRGs.

\begin{figure*}
	\centering
	\includegraphics[width=0.4\linewidth]{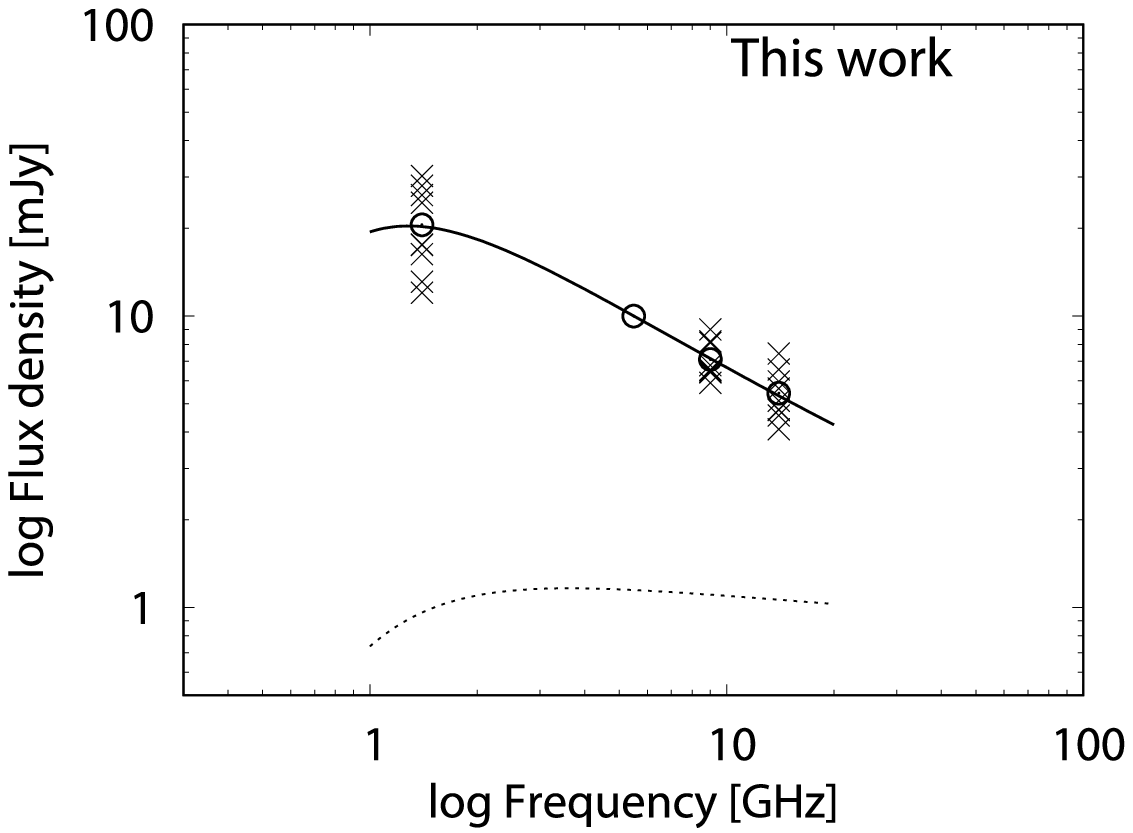}
	\includegraphics[width=0.4\linewidth]{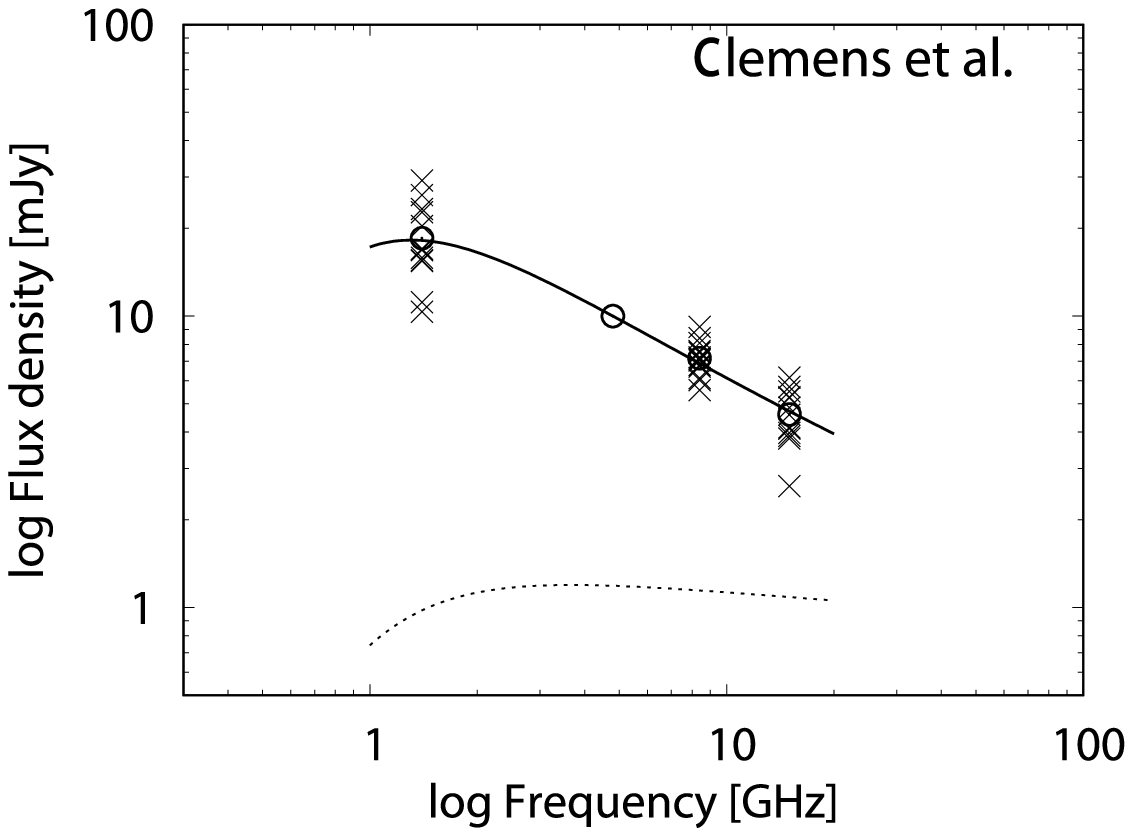}
	\caption{
	    Normalised composite spectra of our targets (left) 
	    and sample from \citet{2008A&A...477...95C} (right).
	    Crosses and open circles show the flux densities of each source and their average flux densities, respectively.
	    We fix flux densities at the C band to 10\,mJy, whose observed frequencies are 5.5 and 4.8\,GHz for our targets and sample from \citet{2008A&A...477...95C}, respectively.
        Solid and dotted lines indicate the best-fit model for the spectra presented by  Equation~\ref{eq:model} and its thermal component, respectively.
        Because of the small thermal fraction, the solid line overlaps the nonthermal component.
	    }
	\label{fig:mean-spectrum}
\end{figure*}

The main results of our observations are as follows: 
The spectral indices at low and high frequencies ($\alpha^{1.4}_{5.0}$ and  $\alpha^{9.0}_{14.0}$) do not relate to each other (Figure~\ref{fig:SPIX-SPIX}).
The FIR-to-radio ratio at 9.0\,GHz, $q_{9.0}$, significantly correlates with the spectral index at high frequency, $\alpha^{9.0}_{14.0}$, while no relation is found at low frequency (Figure~\ref{fig:q-SPIX}).
In contrast, the FIR colour, $f_{60}/f_{100}$, significantly correlates with the spectral index at low frequency, $\alpha^{1.4}_{5.0}$, while no relation is found at high frequency (Figure~\ref{fig:FIR-SPIX}).
These trends are similar to the results obtained by \citet{2008A&A...477...95C}.
For further comparison, composite spectra normalised at flux density in the C band are shown in Figure~\ref{fig:mean-spectrum}, where the model presented in Equation~\ref{eq:model} is also illustrated.
As for data from the previous study, we selected ULIRGs observed at the same bands as our observations (i.e. the L, C, X, and Ku bands).
The overall spectral shapes and the models are comparable between the previous report and ours, and both observations at lower frequencies indicate flux scattering larger than at higher frequencies.

Considering that our targets are selected from \ion{H}{2} and LINER nuclear classifications, the similarities mentioned above imply that radiative processes at radio wavelength might be common in both ULIRGs with and without AGN signatures at optical wavelengths, which is also reported by \citet{2013ApJ...777...58M}.
Since our targets indicate hints of buried AGNs by MIR or submillimetre spectroscopy, this means that radio emission from AGNs could be included in our sample as in the overall ULIRGs sample.
Therefore, the results might confirm the correctness of the previous diagnoses of the buried AGNs \citep{2007ApJS..171...72I,2016AJ....152..218I,2018ApJ...856..143I,2019ApJS..241...19I,2014AJ....148....9I}.

% figure: mean spectra

% -----------------------------------------------------------------------------------
\subsection{Interpretation of statistical trends}\label{sec:sf}
% -----------------------------------------------------------------------------------
In Section~\ref{sec:NIR}, we report that radio spectral indices are not related to the NIR colour index, $(J-K)$, which depends on the dust temperature representing AGN activity  (Figure~\ref{fig:NIR-SPIX}).
This result is interpreted as the absence of statistical contribution of AGN activity to the radio spectra of ULIRGs, which is also reported by \citet{2008A&A...477...95C}.
This is consistent with the prediction by the FIR-to-radio ratio, whose small value compared to other objects is treated as an outlier suggesting the presence of radio-loud AGNs \citep{1991ApJ...378...65C,2001ApJ...554..803Y},
but no such object is found in our sample (Figure~\ref{fig:q-SPIX}).
Although a possible contribution from radio-loud AGNs can be assumed for our sample selected by buried AGNs, no hints of radio-loud AGNs are statistically obtained at this moment.
Hence, our statistical results are primarily explained by other phenomena such as star formation and/or merger activity.

Generally, radio emission from starburst galaxies comes from a mixture of thermal and nonthermal plasma. 
In the radio regime, the former produces FFE and FFA that dominate at higher and lower frequencies, respectively \citep{1991ApJ...378...65C}.
Since both mechanisms share the origin, the correlation of spectral indices can be  assumed between these frequency ranges but have not been found.
One possible explanation is a geometric effect \citep{2008A&A...477...95C}.
For fixed ionising luminosity, the strength of FFE at high frequency (an optically thin regime) does not depend on the size of the ionising source in the centre, whereas the FFA found at low frequency (an optically thick regime) becomes severe if the synchrotron-emitting plasma in the centre is compact.
Therefore, the lack of correlation between $\alpha^{1.4}_{5.0}$ and some physical quantities are easily explained by the fact that the central object has a variety of sizes \citep[cf.][]{2013ApJ...768....2M,2017ApJ...843..117B}.
This picture is consistent with a larger scatter of the flux densities at lower frequencies (Figure~\ref{fig:mean-spectrum}).

Assuming that most radio emission originates from starburst activity even in ULIRGs with a buried AGN (i.e. moderate AGN affects MIR features but not radio emission), both FIR emission and FFE are measures of ongoing starbursts, while synchrotron emission is an indicator of past starbursts that lag behind the lifetime of massive main-sequence stars ($\sim 10^7$\,yrs).
Thus, the FIR-to-radio ratio, $q_{9.0}$, and the ratio of thermal to nonthermal components, $H$, can decrease and the spectral index at high frequency, $\alpha^{9.0}_{14.0}$, can become more negative (i.e. steeper) as the age of the star formation increases.
As discussed in \citet{2008A&A...477...95C}, this scenario explains the relation between $q_{9.0}$ and  $\alpha^{9.0}_{14.0}$ (Figure~\ref{fig:q-SPIX}).

\begin{figure}
	\centering
	\includegraphics[width=0.8\linewidth]{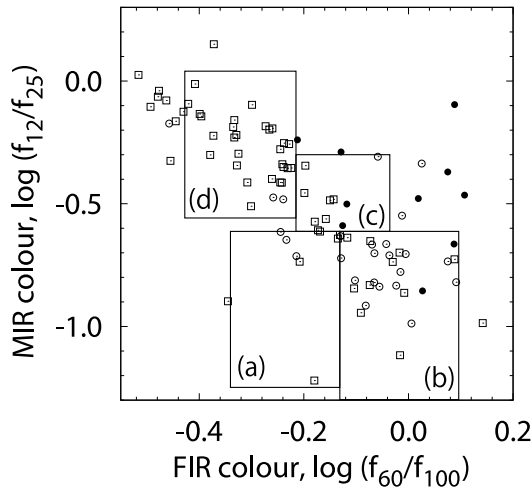}
	\caption{
        IRAS Colour-colour diagrams.
        Horizontal and vertical axes indicate MIR colour $f_{12}/f_{25}$, and FIR colour $f_{60}/f_{100}$, respectively.
        Samples from \citet{2000AJ....120..583D}, \citet{2008A&A...477...95C} and ours are indicated by open squares, open circles, and filled circles, respectively.
        Four closed regions (a)--(d) highlight four main evolutionary phases of the starburst models proposed by \citet{2005MNRAS.364.1286V}, where galaxies transition from (a) to (d).
        }
    \label{fig:MIR-FIR}
\end{figure}

Accepting that the strength of FFA depends on the compactness of the central object,
a correlation between $f_{60}/f_{100}$ and $\alpha^{1.4}_{5.0}$ (Figure~\ref{fig:FIR-SPIX}) suggests that the most compact sources are those in which the dust is the warmest.
As suggested by \citet{2005MNRAS.364.1286V}, the FIR and MIR colours of galaxies change with the age of the starburst. 
Figure~\ref{fig:MIR-FIR} shows their evolution scenario, where the temperature of the dust in galaxies rises as the starburst peaks but then gradually decreases.
This change in FIR colour is not monotonic.
However, because our sample does not include objects belonging to the early starburst phase (a) in the evolutionary model, the dust temperature of the sample is considered to be in the path of the monotonic decrease.
Therefore, if the temperature decreases as the central source becomes large, then the correlation between $f^{60}_{100}$ and $\alpha^{1.4}_{5.0}$ is owing to the evolution.
One explanation is the outward propagation of a ring of star formation \citep{2001ApJ...546..952A,2012MNRAS.420.2209V}.
Note that, at the same time $f^{60}_{100}$ and $\alpha^{1.4}_{5.0}$ do not correlate with $\alpha^{9.0}_{14.0}$ and $q_{9.0}$, respectively, which are measures of the starburst age if we assume a starburst dominant radio spectrum.
Combined with the fact that the size of the star-forming regions may vary from object to object, the age might be only a secondary factor to explain the strength of FFA.

In addition to star formation, merger activity can also affect the radio properties of ULIRGs. \citet{2013ApJ...777...58M} reports relations of the radio properties to merger stages classified by \citet{2011AJ....141..100H}.
As the merger stage progresses, the excess radio emission relative to a prediction from the FIR to radio correlation increases and the radio spectrum at high frequency becomes steeper.
The author suggests that these features found in ongoing merger activity arise from radio bridges and tidal tails similar to what is observed for UGC 12914/5 \citep{1993AJ....105.1730C}.
If this is the case, a correlation between the high-frequency spectral index and the FIR-to-radio ratio due to merger activity is supposed, which is certainly found in Figure~\ref{fig:q-SPIX}.
In contrast, although low-frequency spectra are found to flatten during the merger process, no relation is found between the low-frequency spectral index and the FIR-to-radio ratio.
Because spectral features at lower frequencies are considered to come from the vicinity of the compact starburst \citep{2013ApJ...768....2M}, the lack of the correlation suggests diversity in the optical depth of the FFA screen hiding the central object.

% figure: MIR-FIR plot

% -----------------------------------------------------------------------------------
\subsection{Contribution of AGNs to individual objects}\label{sec:AGN}
% -----------------------------------------------------------------------------------
In Section~\ref{sec:sf}, we find that a hint of AGNs is not suggested by the distribution of the FIR-to-radio ratio where no outlier is found, and by the statistical relation between the NIR colour, $(J-K)$, and radio spectral indices.
The buried AGN found at MIR and submillimetre wavelengths could be considered not to be the main contributors to the radio emission, and hence the radio spectra can be statistically explained by star formation and/or merger activity to some extent \citep[e.g.,][]{2010MNRAS.405..887C,2013ApJ...777...58M}.
However, the situation is much more complicated because $(J-K)$ is not determined by heating from AGNs alone but is also affected by dust extinction, 
and because only galaxies containing radio-loud AGNs have FIR-to-radio ratios distinguishable from those of the starforming galaxies \citep{2009ApJ...706..482M,2010ApJ...724..779M}.
Moreover, at this moment, the factors that determine the radio loudness of AGNs are still not obvious in individual sources.
Therefore, the contribution of another radiative process typically seen in AGNs can still be considered, and the statistical properties and explanations referred to thus far cannot completely rule out the existence of AGNs at radio.

\subsubsection{GHz-peaked spectrum}
Generally, AGNs show strong synchrotron radiation
whose radio spectrum becomes steep if the sources are optically thin 
while peaked or inverted if they suffer from FFA or synchrotron self-absorption 
owing to a strong magnetic field \citep[e.g.][]{2005A&A...435..839T,2008A&A...487..885O}.
In Figure~\ref{fig:spectra}, the nonthermal components of IRAS\,00091$-$0738, IRAS\,01004$-$2237, and IRAS\,01298$-$0744 show a GHz-peaked spectrum.
Although supernova remnants show similar peaked spectra \citep[e.g.][]{1998ApJ...502..218A,2004ApJ...604L..97C}, such a spectrum from AGNs has also been found in ULIRGs \citep[e.g.][]{2004AJ....127..239G,2011AJ....142...17H}, and thus these objects are still probable candidates for AGNs.
Nevertheless, as our observed frequencies are limited to the GHz ranges, the results of the model fit can be different if data at lower frequencies are included \citep[e.g.][]{2004ApJ...604L..97C}.
In reality, many of the spectra, including low-frequency data, indicate the presence of an additional component peaked at frequencies below 1\,GHz  \citep{2010MNRAS.405..887C,2018MNRAS.474..779G}.
To confirm the presence of the GHz-peaked spectrum from the nonthermal component, observations at lower frequencies are required.

Confirmation of a high brightness temperature of more than $10^{10}$\,K found in  sources with synchrotron self-absorption is also important for identifying AGNs \citep{1975ApJ...202..596B}.
The direct measurement of the temperature is the best method, but since the angular resolution of this study is limited, this would be accomplished by future observations with long baselines.
An alternative to high-resolution observation is testing the source variability \citep[e.g.][]{2006ApJ...639..716Z,2007ApJ...661L.139G}.
Although we do not find any significant changes in the sources (Section~\ref{sec:variability}), 
it is insufficient to conclude that the sources are stable because only two-epoch data are available at the moment.

\subsubsection{Steepening at high frequency}
Based on the model given by Equation~\ref{eq:model}, the radio spectrum should flatten at higher frequencies because the contribution of thermal emission showing $\alpha \sim -0.1$ dominates with increasing frequency.
However, in the spectra of IRAS\,00091$-$0738, IRAS\,00188$-$0856, and IRAS\,01298$-$0744, we find steepening toward higher frequencies (see $\alpha^{5.5}_{9.0}$ and $\alpha^{9.0}_{14.0}$ in Table~\ref{tbl:result}).
A similar feature is also reported by several authors,
whose origin has been debated  \citep{2008A&A...477...95C,2010MNRAS.405..887C,2011ApJ...739L..25L,2013ApJ...777...58M}.

One explanation is spectral ageing owing to the energy loss of relativistic electrons, which increases with increasing particle energy, resulting in a change in the spectral slope of $\Delta \alpha \sim 0.5$ across the so-called break frequency \citep{1962SvA.....6..317K,1970ranp.book.....P,1973A&A....26..423J}.
\citet{2008A&A...477...95C} considered that the summation of several components with different break frequencies, in general, produced a straight and monotonic spectrum.
There, stochastic effects such as variations in the supernovae rate might cause the break, but this is not probable because even the brightest supernova remnants contribute only $< 10$\% of the total emission from ULIRG's nuclei \citep[e.g.][]{2006ApJ...647..185L}.
However, synchrotron ageing can still explain the spectral break if we consider nonthermal plasma from AGNs whose contribution becomes potentially well over 10\%.
In particular, large nonthermal fractions (Figure~\ref{fig:spectra}) are found in IRAS\,00091$-$0738 and IRAS\,00188$-$0856, which are probable candidates for sources with radio AGNs.
Although at $\sim$\,30\,GHz the presence of a predominant thermal component is found for typical ULIRGs \citep{2017ApJ...843..117B}, the AGN candidates reported here might suggest a significant contribution of a nonthermal component at the frequency.
Therefore, as suggested by \citet{2011ApJ...739L..25L}, the detection of spectral breaks can still be a way to distinguish AGNs from starbursts.
We note that, while an instantaneous injection model is preferred if multifrequency images with the sufficiently high spatial resolution are available \citep[e.g.][]{2006ApJ...648..148N}, a continuous injection model is supposed for unresolved sources, whose break does not disappear even if several components of different spectral ages are superposed \citep[e.g.][]{1999A&A...345..769M,2003PASA...20...19M}.

In addition to the above, some authors have made other arguments.
\citet{2010MNRAS.405..887C} has proposed that the ionisation loss that gradually flattens radio spectra at lower frequencies explained the spectral break \citep{2006ApJ...645..186T}.
Nevertheless, some of their sample at the same time show a highly steep spectral index of $\alpha < -1$ at high frequency \citep{2008A&A...477...95C}, which still requires the synchrotron ageing of nonthermal plasma as found in AGNs.
On another hand, a consideration made by \citet{2013ApJ...777...58M} has not only explained the spectral break but also the highly steep spectral index.
As referred in Section \ref{sec:sf}, they have reported that as the merger stage of merging galaxy pairs progresses radio bridges and tidal tails of ongoing mergers start to show a steep spectral index at high frequency, which is caused by rapid synchrotron ageing in a magnetised medium produced by the dynamical interaction.
This scenario could be the case for IRAS\,00091$-$0738 and IRAS\,01298$-$0744, whose tidal tails are optically confirmed by \citet{2002ApJS..143..315V}.
In contrast, the optical image of IRAS\,00188$-$0856 provided by the same authors does not show any signs of tidal tails or disturbed morphology.
In contrast, the optical image of IRAS\,00188$-$0856 provided by the same authors does not show any signs of tidal tails or disturbed morphology.
We do not find any traces of merger activity at least in the optical image and thus another explanation like an AGN is still considerable for IRAS\,00188$-$0856.

% -----------------------------------------------------------------------------------
\subsubsection{Extended emission in IRAS\,01004--2237}\label{sec:01004}
% -----------------------------------------------------------------------------------
IRAS\,01004$-$2237 shows extended features across $\sim 50$\,arcsec, which corresponds to the projected linear size of $\sim100$\,kpc (Figure~\ref{fig:im-IRAS01004}).
A few arcseconds away from the northeast component, we find a faint point source in an optical image provided by the Pan-STARRS \citep{2016arXiv161205560C} whose colour suggests a galaxy at $0.6<z<1.0$ \citep[cf.][]{2015ApJ...803..105P}, while the southwest one has no counterpart, which should be associated with IRAS\,01004$-$2237 itself.
Its symmetrical structure including the northeast component evokes FR2 radio galaxies \citep{1974MNRAS.167P..31F}.

While most ULIRGs generally constitute merger systems, IRAS\,01004$-$2237 itself does not indicate the implication of recent merger activity in the form of tidal tails or loops \citep{2002ApJS..143..315V}.
In addition, although its luminosity does not meet the criteria, the source has a nearly stellar radial profile similar to that of quasars \citep{1998ApJ...492..116S}.
Furthermore, \citet{1999ApJ...522..113V} classified the source as \ion{H}{2}  once, but \citet{2010ApJ...709..884Y} and \citet{2013MNRAS.432..138R} identify the optical AGN signature by decomposing its emission lines into multiple components.
Combined with the diagnostic result at MIR \citep{2007ApJS..171...72I,2009ApJS..182..628V,2010MNRAS.405.2505N}, the presence of an AGN in the nucleus of IRAS\,01004$-$2237 is certain, which supports the AGN origin of the extended radio emission.

On the other hand, the extended emission could also be considered as a radio mini-halo or a radio relic associated with galaxy clusters \citep{2012A&ARv..20...54F,2019SSRv..215...16V}.
These are usually located in the centre of merging clusters and are accompanied by a hot ionised intracluster medium that emits X-rays with the same spread as radio.
However, we do not found such a feature by X-ray observations  \citep{1999A&A...349..389V,2005ApJ...633..664T}; hence, the extended emission is not induced by the merger.

Regardless of whether the extended feature originates from an AGN or a merger activity, an interpretation of the radio spectrum (Figure~\ref{fig:spectra_IRAS01004_extended}) is still unknown.
At high frequency, nonthermal plasma having lost the energy generally shows a steep spectrum indicating $\alpha\lesssim-1$, or up to $\sim -2$ \citep{2007A&A...470..875P,2008Natur.455..944B,2010MNRAS.402.1892O,2011A&A...526A.148M}.
By contrast, the spectrum of the extended component associated with IRAS\,01004--2237 shows flattening at high frequency, which suggests a significant contribution of the thermal plasma.
However, this kind of feature has not been reported for both AGN or merger activities.
We must pay attention to the interpretation of this data because this result could be an artificial one.
Since the spatial resolution of the images at 5.5 and 14.0\,GHz is approximately two times lower than that at 9.0\,GHz (Table\,\ref{tbl:imaging}), our observation can resolve out the diffuse extended feature at 9.0\,GHz and then underestimate the flux density.
We need deep observation to obtain a conclusive result before making further discussions.

% -----------------------------------------------------------------------------------
\section{Conclusions}\label{sec:conlusion}
% -----------------------------------------------------------------------------------
We conducted new multifrequency VLA observations for ULIRGs whose AGNs are not found at optical wavelengths but are identified at MIR or submillimetre wavelengths \citep{2007ApJS..171...72I,2016AJ....152..218I,2018ApJ...856..143I,2019ApJS..241...19I,2014AJ....148....9I}.
This was made for shedding light on the contribution of buried AGNs to radio emission of ULIRGs through a comparison with the observational results obtained by \citet{2008A&A...477...95C}, whose sample is based on the entire ULIRGs.
The main conclusions of this study can be summarised as follows.
\begin{enumerate}
    \item 
The spectral indices at low and high frequencies are not related to each other.
The spectral index at high frequency significantly correlates with the FIR-to-radio ratio at 9.0\,GHz while no relation is found at low frequency.
In contrast, the spectral index at low frequency significantly correlates with the FIR colour while no relation is found at high frequency.
These statistical trends are similar to those of the entire ULIRGs provided by \citet{2008A&A...477...95C}.
    \item 
The statistical similarity mentioned above suggests that all ULIRGs have common radiative processes at radio regardless of the presence of optical AGNs.
Thus, whereas the radio spectra of our targets can be mainly explained by starbursts and/or merger activity \citep{2010MNRAS.405..887C,2013ApJ...777...58M}, combined with the AGN diagnosis by MIR and submillimetre spectroscopy it might be possible that AGNs contribute equally to the radio emissions of every ULIRG.
    \item
Some individual sources imply contributions from AGNs.
IRAS\,00188$-$0856 not indicating disturbed morphology at optical wavelengths shows a high nonthermal fraction and a spectral break at high frequency, which can be explained by synchrotron ageing of nonthermal plasma emitted from AGNs.
In addition, 100-kpc scale extended radio emission is associated with IRAS\,01004$-$2237, whose two-sided morphology and no X-ray detection suggest that this system is not induced by a merger in a cluster but originates from AGN activity.
\end{enumerate}

% -----------------------------------------------------------------------------------
% Acknowledgement
% -----------------------------------------------------------------------------------
\section*{Acknowledgement}
We wish to express our appreciation to an anonymous reviewer for the insightful comments helping us significantly improve the paper and make a solid discussion.
The VLA is a facility of NRAO, operated by Associated University Inc. under cooperative agreement with the National Science Foundation.
This publication makes use of data products from the Two Micron All Sky Survey, which is a joint project of the University of Massachusetts and the Infrared Processing and Analysis Center/California Institute of Technology, funded by the National Aeronautics and Space Administration and the National Science Foundation.
We also made use of the NASA/IPAC Extragalactic Database operated by the Jet Propulsion Laboratory, California Institute of Technology.
This work was partially supported by a Grant-in-Aid for Scientific Research (B; 24340042) from the Japanese Ministry of Education, Culture, Sports, Science and Technology.
We thank Editage (www.editage.com) for English language editing.

\section*{Data Availability}
Data underlying this article are available in NRAO Science Data Archive, at 
\href{https://archive.nrao.edu/archive/advquery.jsp}
{https://archive.nrao.edu/archive/advquery.jsp}. 
The reduced data by authors will be shared on reasonable request to the corresponding author.

\bibliographystyle{mnras}
\bibliography{ULIRG_2020.bib}

\begin{thebibliography}{}
\makeatletter
\relax
\def\mn@urlcharsother{\let\do\@makeother \do\$\do\&\do\#\do\^\do\_\do\%\do\~}
\def\mn@doi{\begingroup\mn@urlcharsother \@ifnextchar [ {\mn@doi@}
  {\mn@doi@[]}}
\def\mn@doi@[#1]#2{\def\@tempa{#1}\ifx\@tempa\@empty \href
  {http://dx.doi.org/#2} {doi:#2}\else \href {http://dx.doi.org/#2} {#1}\fi
  \endgroup}
\def\mn@eprint#1#2{\mn@eprint@#1:#2::\@nil}
\def\mn@eprint@arXiv#1{\href {http://arxiv.org/abs/#1} {{\tt arXiv:#1}}}
\def\mn@eprint@dblp#1{\href {http://dblp.uni-trier.de/rec/bibtex/#1.xml}
  {dblp:#1}}
\def\mn@eprint@#1:#2:#3:#4\@nil{\def\@tempa {#1}\def\@tempb {#2}\def\@tempc
  {#3}\ifx \@tempc \@empty \let \@tempc \@tempb \let \@tempb \@tempa \fi \ifx
  \@tempb \@empty \def\@tempb {arXiv}\fi \@ifundefined
  {mn@eprint@\@tempb}{\@tempb:\@tempc}{\expandafter \expandafter \csname
  mn@eprint@\@tempb\endcsname \expandafter{\@tempc}}}

\bibitem[\protect\citeauthoryear{{Allen} \& {Kronberg}}{{Allen} \&
  {Kronberg}}{1998}]{1998ApJ...502..218A}
{Allen} M.~L.,  {Kronberg} P.~P.,  1998, \mn@doi [\apj] {10.1086/305894}, \href
  {http://adsabs.harvard.edu/abs/1998ApJ...502..218A} {502, 218}

\bibitem[\protect\citeauthoryear{{Alonso-Herrero}, {Engelbracht}, {Rieke},
  {Rieke}  \& {Quillen}}{{Alonso-Herrero} et~al.}{2001}]{2001ApJ...546..952A}
{Alonso-Herrero} A.,  {Engelbracht} C.~W.,  {Rieke} M.~J.,  {Rieke} G.~H.,
  {Quillen} A.~C.,  2001, \mn@doi [\apj] {10.1086/318282}, \href
  {https://ui.adsabs.harvard.edu/abs/2001ApJ...546..952A} {546, 952}

\bibitem[\protect\citeauthoryear{{Armus} et~al.,}{{Armus}
  et~al.}{2004}]{2004ApJS..154..178A}
{Armus} L.,  et~al., 2004, \mn@doi [\apjs] {10.1086/422915}, \href
  {https://ui.adsabs.harvard.edu/abs/2004ApJS..154..178A} {154, 178}

\bibitem[\protect\citeauthoryear{{Armus} et~al.,}{{Armus}
  et~al.}{2007}]{2007ApJ...656..148A}
{Armus} L.,  et~al., 2007, \mn@doi [\apj] {10.1086/510107}, \href
  {http://ads.nao.ac.jp/abs/2007ApJ...656..148A} {656, 148}

\bibitem[\protect\citeauthoryear{{Barcos-Mu{\~n}oz} et~al.,}{{Barcos-Mu{\~n}oz}
  et~al.}{2017}]{2017ApJ...843..117B}
{Barcos-Mu{\~n}oz} L.,  et~al., 2017, \mn@doi [\apj]
  {10.3847/1538-4357/aa789a}, \href
  {https://ui.adsabs.harvard.edu/abs/2017ApJ...843..117B} {843, 117}

\bibitem[\protect\citeauthoryear{{Becker}, {White}  \& {Helfand}}{{Becker}
  et~al.}{1995}]{1995ApJ...450..559B}
{Becker} R.~H.,  {White} R.~L.,   {Helfand} D.~J.,  1995, \mn@doi [\apj]
  {10.1086/176166}, \href
  {https://ui.adsabs.harvard.edu/abs/1995ApJ...450..559B} {450, 559}

\bibitem[\protect\citeauthoryear{{Broderick} \& {Condon}}{{Broderick} \&
  {Condon}}{1975}]{1975ApJ...202..596B}
{Broderick} J.~J.,  {Condon} J.~J.,  1975, \mn@doi [\apj] {10.1086/154012},
  \href {https://ui.adsabs.harvard.edu/abs/1975ApJ...202..596B} {202, 596}

\bibitem[\protect\citeauthoryear{{Brunetti} et~al.,}{{Brunetti}
  et~al.}{2008}]{2008Natur.455..944B}
{Brunetti} G.,  et~al., 2008, \mn@doi [\nat] {10.1038/nature07379}, \href
  {https://ui.adsabs.harvard.edu/abs/2008Natur.455..944B} {455, 944}

\bibitem[\protect\citeauthoryear{{Chambers} et~al.,}{{Chambers}
  et~al.}{2016}]{2016arXiv161205560C}
{Chambers} K.~C.,  et~al., 2016, arXiv e-prints, \href
  {https://ui.adsabs.harvard.edu/abs/2016arXiv161205560C} {p. arXiv:1612.05560}

\bibitem[\protect\citeauthoryear{{Chandra}, {Ray}  \& {Bhatnagar}}{{Chandra}
  et~al.}{2004}]{2004ApJ...604L..97C}
{Chandra} P.,  {Ray} A.,   {Bhatnagar} S.,  2004, \mn@doi [\apjl]
  {10.1086/383615}, \href {http://ads.nao.ac.jp/abs/2004ApJ...604L..97C} {604,
  L97}

\bibitem[\protect\citeauthoryear{{Clemens}, {Vega}, {Bressan}, {Granato},
  {Silva}  \& {Panuzzo}}{{Clemens} et~al.}{2008}]{2008A&A...477...95C}
{Clemens} M.~S.,  {Vega} O.,  {Bressan} A.,  {Granato} G.~L.,  {Silva} L.,
  {Panuzzo} P.,  2008, \mn@doi [\aap] {10.1051/0004-6361:20077224}, \href
  {https://ui.adsabs.harvard.edu/abs/2008A&A...477...95C} {477, 95}

\bibitem[\protect\citeauthoryear{{Clemens}, {Scaife}, {Vega}  \&
  {Bressan}}{{Clemens} et~al.}{2010}]{2010MNRAS.405..887C}
{Clemens} M.~S.,  {Scaife} A.,  {Vega} O.,   {Bressan} A.,  2010, \mn@doi
  [\mnras] {10.1111/j.1365-2966.2010.16534.x}, \href
  {https://ui.adsabs.harvard.edu/abs/2010MNRAS.405..887C} {405, 887}

\bibitem[\protect\citeauthoryear{{Condon}, {Huang}, {Yin}  \& {Thuan}}{{Condon}
  et~al.}{1991}]{1991ApJ...378...65C}
{Condon} J.~J.,  {Huang} Z.~P.,  {Yin} Q.~F.,   {Thuan} T.~X.,  1991, \mn@doi
  [\apj] {10.1086/170407}, \href
  {https://ui.adsabs.harvard.edu/abs/1991ApJ...378...65C} {378, 65}

\bibitem[\protect\citeauthoryear{{Condon}, {Helou}, {Sanders}  \&
  {Soifer}}{{Condon} et~al.}{1993}]{1993AJ....105.1730C}
{Condon} J.~J.,  {Helou} G.,  {Sanders} D.~B.,   {Soifer} B.~T.,  1993, \mn@doi
  [\aj] {10.1086/116549}, \href
  {https://ui.adsabs.harvard.edu/abs/1993AJ....105.1730C} {105, 1730}

\bibitem[\protect\citeauthoryear{{Condon}, {Cotton}, {Greisen}, {Yin},
  {Perley}, {Taylor}  \& {Broderick}}{{Condon}
  et~al.}{1998}]{1998AJ....115.1693C}
{Condon} J.~J.,  {Cotton} W.~D.,  {Greisen} E.~W.,  {Yin} Q.~F.,  {Perley}
  R.~A.,  {Taylor} G.~B.,   {Broderick} J.~J.,  1998, \mn@doi [\aj]
  {10.1086/300337}, \href
  {https://ui.adsabs.harvard.edu/abs/1998AJ....115.1693C} {115, 1693}

\bibitem[\protect\citeauthoryear{{Cutri} et~al.,}{{Cutri}
  et~al.}{2003}]{2003yCat.2246....0C}
{Cutri} R.~M.,  et~al., 2003, VizieR Online Data Catalog, \href
  {https://ui.adsabs.harvard.edu/abs/2003yCat.2246....0C} {p. II/246}

\bibitem[\protect\citeauthoryear{{Dale} et~al.,}{{Dale}
  et~al.}{2000}]{2000AJ....120..583D}
{Dale} D.~A.,  et~al., 2000, \mn@doi [\aj] {10.1086/301478}, \href
  {https://ui.adsabs.harvard.edu/abs/2000AJ....120..583D} {120, 583}

\bibitem[\protect\citeauthoryear{{Fanaroff} \& {Riley}}{{Fanaroff} \&
  {Riley}}{1974}]{1974MNRAS.167P..31F}
{Fanaroff} B.~L.,  {Riley} J.~M.,  1974, \mn@doi [\mnras]
  {10.1093/mnras/167.1.31P}, \href
  {https://ui.adsabs.harvard.edu/abs/1974MNRAS.167P..31F} {167, 31P}

\bibitem[\protect\citeauthoryear{{Feretti}, {Giovannini}, {Govoni}  \&
  {Murgia}}{{Feretti} et~al.}{2012}]{2012A&ARv..20...54F}
{Feretti} L.,  {Giovannini} G.,  {Govoni} F.,   {Murgia} M.,  2012, \mn@doi
  [\aapr] {10.1007/s00159-012-0054-z}, \href
  {https://ui.adsabs.harvard.edu/abs/2012A&ARv..20...54F} {20, 54}

\bibitem[\protect\citeauthoryear{{Gallimore} \& {Beswick}}{{Gallimore} \&
  {Beswick}}{2004}]{2004AJ....127..239G}
{Gallimore} J.~F.,  {Beswick} R.,  2004, \mn@doi [\aj] {10.1086/379959}, \href
  {https://ui.adsabs.harvard.edu/abs/2004AJ....127..239G} {127, 239}

\bibitem[\protect\citeauthoryear{{Galvin} et~al.,}{{Galvin}
  et~al.}{2018}]{2018MNRAS.474..779G}
{Galvin} T.~J.,  et~al., 2018, \mn@doi [\mnras] {10.1093/mnras/stx2613}, \href
  {https://ui.adsabs.harvard.edu/abs/2018MNRAS.474..779G} {474, 779}

\bibitem[\protect\citeauthoryear{{Ghosh} \& {Punsly}}{{Ghosh} \&
  {Punsly}}{2007}]{2007ApJ...661L.139G}
{Ghosh} K.~K.,  {Punsly} B.,  2007, \mn@doi [\apjl] {10.1086/518859}, \href
  {https://ui.adsabs.harvard.edu/abs/2007ApJ...661L.139G} {661, L139}

\bibitem[\protect\citeauthoryear{{Haan} et~al.,}{{Haan}
  et~al.}{2011}]{2011AJ....141..100H}
{Haan} S.,  et~al., 2011, \mn@doi [\aj] {10.1088/0004-6256/141/3/100}, \href
  {https://ui.adsabs.harvard.edu/abs/2011AJ....141..100H} {141, 100}

\bibitem[\protect\citeauthoryear{{Hagiwara}, {Baan}  \&
  {Kl{\"o}ckner}}{{Hagiwara} et~al.}{2011}]{2011AJ....142...17H}
{Hagiwara} Y.,  {Baan} W.~A.,   {Kl{\"o}ckner} H.-R.,  2011, \mn@doi [\aj]
  {10.1088/0004-6256/142/1/17}, \href
  {https://ui.adsabs.harvard.edu/abs/2011AJ....142...17H} {142, 17}

\bibitem[\protect\citeauthoryear{{Helou}, {Soifer}  \&
  {Rowan-Robinson}}{{Helou} et~al.}{1985}]{1985ApJ...298L...7H}
{Helou} G.,  {Soifer} B.~T.,   {Rowan-Robinson} M.,  1985, \mn@doi [\apjl]
  {10.1086/184556}, \href
  {https://ui.adsabs.harvard.edu/abs/1985ApJ...298L...7H} {298, L7}

\bibitem[\protect\citeauthoryear{{Imanishi}}{{Imanishi}}{2009}]{2009ApJ...694..751I}
{Imanishi} M.,  2009, \mn@doi [\apj] {10.1088/0004-637X/694/2/751}, \href
  {http://ads.nao.ac.jp/abs/2009ApJ...694..751I} {694, 751}

\bibitem[\protect\citeauthoryear{{Imanishi} \& {Nakanishi}}{{Imanishi} \&
  {Nakanishi}}{2014}]{2014AJ....148....9I}
{Imanishi} M.,  {Nakanishi} K.,  2014, \mn@doi [\aj]
  {10.1088/0004-6256/148/1/9}, \href
  {https://ui.adsabs.harvard.edu/abs/2014AJ....148....9I} {148, 9}

\bibitem[\protect\citeauthoryear{{Imanishi}, {Dudley}  \& {Maloney}}{{Imanishi}
  et~al.}{2006}]{2006ApJ...637..114I}
{Imanishi} M.,  {Dudley} C.~C.,   {Maloney} P.~R.,  2006, \mn@doi [\apj]
  {10.1086/498391}, \href
  {https://ui.adsabs.harvard.edu/abs/2006ApJ...637..114I} {637, 114}

\bibitem[\protect\citeauthoryear{{Imanishi}, {Dudley}, {Maiolino}, {Maloney},
  {Nakagawa}  \& {Risaliti}}{{Imanishi} et~al.}{2007}]{2007ApJS..171...72I}
{Imanishi} M.,  {Dudley} C.~C.,  {Maiolino} R.,  {Maloney} P.~R.,  {Nakagawa}
  T.,   {Risaliti} G.,  2007, \mn@doi [\apjs] {10.1086/513715}, \href
  {https://ui.adsabs.harvard.edu/abs/2007ApJS..171...72I} {171, 72}

\bibitem[\protect\citeauthoryear{{Imanishi}, {Nakagawa}, {Ohyama}, {Shirahata},
  {Wada}, {Onaka}  \& {Oi}}{{Imanishi} et~al.}{2008}]{2008PASJ...60S.489I}
{Imanishi} M.,  {Nakagawa} T.,  {Ohyama} Y.,  {Shirahata} M.,  {Wada} T.,
  {Onaka} T.,   {Oi} N.,  2008, \pasj, \href
  {http://ads.nao.ac.jp/abs/2008PASJ...60S.489I} {60, 489}

\bibitem[\protect\citeauthoryear{{Imanishi}, {Maiolino}  \&
  {Nakagawa}}{{Imanishi} et~al.}{2010a}]{2010ApJ...709..801I}
{Imanishi} M.,  {Maiolino} R.,   {Nakagawa} T.,  2010a, \mn@doi [\apj]
  {10.1088/0004-637X/709/2/801}, \href
  {http://ads.nao.ac.jp/abs/2010ApJ...709..801I} {709, 801}

\bibitem[\protect\citeauthoryear{{Imanishi}, {Nakagawa}, {Shirahata}, {Ohyama}
  \& {Onaka}}{{Imanishi} et~al.}{2010b}]{2010ApJ...721.1233I}
{Imanishi} M.,  {Nakagawa} T.,  {Shirahata} M.,  {Ohyama} Y.,   {Onaka} T.,
  2010b, \mn@doi [\apj] {10.1088/0004-637X/721/2/1233}, \href
  {http://ads.nao.ac.jp/abs/2010ApJ...721.1233I} {721, 1233}

\bibitem[\protect\citeauthoryear{{Imanishi}, {Nakanishi}  \&
  {Izumi}}{{Imanishi} et~al.}{2016}]{2016AJ....152..218I}
{Imanishi} M.,  {Nakanishi} K.,   {Izumi} T.,  2016, \mn@doi [\aj]
  {10.3847/0004-6256/152/6/218}, \href
  {https://ui.adsabs.harvard.edu/abs/2016AJ....152..218I} {152, 218}

\bibitem[\protect\citeauthoryear{{Imanishi}, {Nakanishi}  \&
  {Izumi}}{{Imanishi} et~al.}{2018}]{2018ApJ...856..143I}
{Imanishi} M.,  {Nakanishi} K.,   {Izumi} T.,  2018, \mn@doi [\apj]
  {10.3847/1538-4357/aab42f}, \href
  {https://ui.adsabs.harvard.edu/abs/2018ApJ...856..143I} {856, 143}

\bibitem[\protect\citeauthoryear{{Imanishi}, {Nakanishi}  \&
  {Izumi}}{{Imanishi} et~al.}{2019}]{2019ApJS..241...19I}
{Imanishi} M.,  {Nakanishi} K.,   {Izumi} T.,  2019, \mn@doi [\apjs]
  {10.3847/1538-4365/ab05b9}, \href
  {https://ui.adsabs.harvard.edu/abs/2019ApJS..241...19I} {241, 19}

\bibitem[\protect\citeauthoryear{{Jaffe} \& {Perola}}{{Jaffe} \&
  {Perola}}{1973}]{1973A&A....26..423J}
{Jaffe} W.~J.,  {Perola} G.~C.,  1973, \aap, \href
  {https://ui.adsabs.harvard.edu/abs/1973A&A....26..423J} {26, 423}

\bibitem[\protect\citeauthoryear{{Kardashev}}{{Kardashev}}{1962}]{1962SvA.....6..317K}
{Kardashev} N.~S.,  1962, \sovast, \href
  {https://ui.adsabs.harvard.edu/abs/1962SvA.....6..317K} {6, 317}

\bibitem[\protect\citeauthoryear{{Kauffmann} et~al.,}{{Kauffmann}
  et~al.}{2003}]{2003MNRAS.346.1055K}
{Kauffmann} G.,  et~al., 2003, \mn@doi [\mnras]
  {10.1111/j.1365-2966.2003.07154.x}, \href
  {https://ui.adsabs.harvard.edu/abs/2003MNRAS.346.1055K} {346, 1055}

\bibitem[\protect\citeauthoryear{{Kewley}, {Heisler}, {Dopita}  \&
  {Lumsden}}{{Kewley} et~al.}{2001}]{2001ApJS..132...37K}
{Kewley} L.~J.,  {Heisler} C.~A.,  {Dopita} M.~A.,   {Lumsden} S.,  2001,
  \mn@doi [\apjs] {10.1086/318944}, \href
  {http://ads.nao.ac.jp/abs/2001ApJS..132...37K} {132, 37}

\bibitem[\protect\citeauthoryear{{Kim} \& {Sanders}}{{Kim} \&
  {Sanders}}{1998}]{1998ApJS..119...41K}
{Kim} D.~C.,  {Sanders} D.~B.,  1998, \mn@doi [\apjs] {10.1086/313148}, \href
  {https://ui.adsabs.harvard.edu/abs/1998ApJS..119...41K} {119, 41}

\bibitem[\protect\citeauthoryear{{Leroy} et~al.,}{{Leroy}
  et~al.}{2011}]{2011ApJ...739L..25L}
{Leroy} A.~K.,  et~al., 2011, \mn@doi [\apjl] {10.1088/2041-8205/739/1/L25},
  \href {https://ui.adsabs.harvard.edu/abs/2011ApJ...739L..25L} {739, L25}

\bibitem[\protect\citeauthoryear{{Lister}}{{Lister}}{2001}]{2001ApJ...561..676L}
{Lister} M.~L.,  2001, \mn@doi [\apj] {10.1086/323528}, \href
  {http://ads.nao.ac.jp/abs/2001ApJ...561..676L} {561, 676}

\bibitem[\protect\citeauthoryear{{Lonsdale}, {Diamond}, {Thrall}, {Smith}  \&
  {Lonsdale}}{{Lonsdale} et~al.}{2006}]{2006ApJ...647..185L}
{Lonsdale} C.~J.,  {Diamond} P.~J.,  {Thrall} H.,  {Smith} H.~E.,   {Lonsdale}
  C.~J.,  2006, \mn@doi [\apj] {10.1086/505193}, \href
  {https://ui.adsabs.harvard.edu/abs/2006ApJ...647..185L} {647, 185}

\bibitem[\protect\citeauthoryear{{Maiolino} et~al.,}{{Maiolino}
  et~al.}{2003}]{2003MNRAS.344L..59M}
{Maiolino} R.,  et~al., 2003, \mn@doi [\mnras]
  {10.1046/j.1365-8711.2003.07036.x}, \href
  {https://ui.adsabs.harvard.edu/abs/2003MNRAS.344L..59M} {344, L59}

\bibitem[\protect\citeauthoryear{{McMullin}, {Waters}, {Schiebel}, {Young}  \&
  {Golap}}{{McMullin} et~al.}{2007}]{2007ASPC..376..127M}
{McMullin} J.~P.,  {Waters} B.,  {Schiebel} D.,  {Young} W.,   {Golap} K.,
  2007, in {Shaw} R.~A.,  {Hill} F.,   {Bell} D.~J.,  eds,  Astronomical
  Society of the Pacific Conference Series Vol. 376, Astronomical Data Analysis
  Software and Systems XVI. p.~127

\bibitem[\protect\citeauthoryear{{Mori{\'c}}, {Smol{\v{c}}i{\'c}}, {Kimball},
  {Riechers}, {Ivezi{\'c}}  \& {Scoville}}{{Mori{\'c}}
  et~al.}{2010}]{2010ApJ...724..779M}
{Mori{\'c}} I.,  {Smol{\v{c}}i{\'c}} V.,  {Kimball} A.,  {Riechers} D.~A.,
  {Ivezi{\'c}} {\v{Z}}.,   {Scoville} N.,  2010, \mn@doi [\apj]
  {10.1088/0004-637X/724/1/779}, \href
  {https://ui.adsabs.harvard.edu/abs/2010ApJ...724..779M} {724, 779}

\bibitem[\protect\citeauthoryear{{Moshir} \& {et al.}}{{Moshir} \& {et
  al.}}{1990}]{1990IRASF.C......0M}
{Moshir} M.,  {et al.} 1990, IRAS Faint Source Catalogue, \href
  {https://ui.adsabs.harvard.edu/abs/1990IRASF.C......0M} {p.~0}

\bibitem[\protect\citeauthoryear{{Murgia}}{{Murgia}}{2003}]{2003PASA...20...19M}
{Murgia} M.,  2003, \mn@doi [\pasa] {10.1071/AS02033}, \href
  {http://ads.nao.ac.jp/abs/2003PASA...20...19M} {20, 19}

\bibitem[\protect\citeauthoryear{{Murgia}, {Fanti}, {Fanti}, {Gregorini},
  {Klein}, {Mack}  \& {Vigotti}}{{Murgia} et~al.}{1999}]{1999A&A...345..769M}
{Murgia} M.,  {Fanti} C.,  {Fanti} R.,  {Gregorini} L.,  {Klein} U.,  {Mack}
  K.~H.,   {Vigotti} M.,  1999, \aap, \href
  {https://ui.adsabs.harvard.edu/abs/1999A&A...345..769M} {345, 769}

\bibitem[\protect\citeauthoryear{{Murgia} et~al.,}{{Murgia}
  et~al.}{2011}]{2011A&A...526A.148M}
{Murgia} M.,  et~al., 2011, \mn@doi [\aap] {10.1051/0004-6361/201015302}, \href
  {https://ui.adsabs.harvard.edu/abs/2011A&A...526A.148M} {526, A148}

\bibitem[\protect\citeauthoryear{{Murphy}}{{Murphy}}{2009}]{2009ApJ...706..482M}
{Murphy} E.~J.,  2009, \mn@doi [\apj] {10.1088/0004-637X/706/1/482}, \href
  {https://ui.adsabs.harvard.edu/abs/2009ApJ...706..482M} {706, 482}

\bibitem[\protect\citeauthoryear{{Murphy}}{{Murphy}}{2013}]{2013ApJ...777...58M}
{Murphy} E.~J.,  2013, \mn@doi [\apj] {10.1088/0004-637X/777/1/58}, \href
  {https://ui.adsabs.harvard.edu/abs/2013ApJ...777...58M} {777, 58}

\bibitem[\protect\citeauthoryear{{Murphy}, {Stierwalt}, {Armus}, {Condon}  \&
  {Evans}}{{Murphy} et~al.}{2013}]{2013ApJ...768....2M}
{Murphy} E.~J.,  {Stierwalt} S.,  {Armus} L.,  {Condon} J.~J.,   {Evans} A.~S.,
   2013, \mn@doi [\apj] {10.1088/0004-637X/768/1/2}, \href
  {https://ui.adsabs.harvard.edu/abs/2013ApJ...768....2M} {768, 2}

\bibitem[\protect\citeauthoryear{{Nagai}, {Inoue}, {Asada}, {Kameno}  \&
  {Doi}}{{Nagai} et~al.}{2006}]{2006ApJ...648..148N}
{Nagai} H.,  {Inoue} M.,  {Asada} K.,  {Kameno} S.,   {Doi} A.,  2006, \mn@doi
  [\apj] {10.1086/505793}, \href
  {https://ui.adsabs.harvard.edu/abs/2006ApJ...648..148N} {648, 148}

\bibitem[\protect\citeauthoryear{{Nardini}, {Risaliti}, {Salvati}, {Sani},
  {Imanishi}, {Marconi}  \& {Maiolino}}{{Nardini}
  et~al.}{2008}]{2008MNRAS.385L.130N}
{Nardini} E.,  {Risaliti} G.,  {Salvati} M.,  {Sani} E.,  {Imanishi} M.,
  {Marconi} A.,   {Maiolino} R.,  2008, \mn@doi [\mnras]
  {10.1111/j.1745-3933.2008.00450.x}, \href
  {http://ads.nao.ac.jp/abs/2008MNRAS.385L.130N} {385, L130}

\bibitem[\protect\citeauthoryear{{Nardini}, {Risaliti}, {Salvati}, {Sani},
  {Watabe}, {Marconi}  \& {Maiolino}}{{Nardini}
  et~al.}{2009}]{2009MNRAS.399.1373N}
{Nardini} E.,  {Risaliti} G.,  {Salvati} M.,  {Sani} E.,  {Watabe} Y.,
  {Marconi} A.,   {Maiolino} R.,  2009, \mn@doi [\mnras]
  {10.1111/j.1365-2966.2009.15357.x}, \href
  {http://ads.nao.ac.jp/abs/2009MNRAS.399.1373N} {399, 1373}

\bibitem[\protect\citeauthoryear{{Nardini}, {Risaliti}, {Watabe}, {Salvati}  \&
  {Sani}}{{Nardini} et~al.}{2010}]{2010MNRAS.405.2505N}
{Nardini} E.,  {Risaliti} G.,  {Watabe} Y.,  {Salvati} M.,   {Sani} E.,  2010,
  \mn@doi [\mnras] {10.1111/j.1365-2966.2010.16618.x}, \href
  {http://ads.nao.ac.jp/abs/2010MNRAS.405.2505N} {405, 2505}

\bibitem[\protect\citeauthoryear{{Orienti} \& {Dallacasa}}{{Orienti} \&
  {Dallacasa}}{2008}]{2008A&A...487..885O}
{Orienti} M.,  {Dallacasa} D.,  2008, \mn@doi [\aap]
  {10.1051/0004-6361:200809948}, \href
  {http://adsabs.harvard.edu/abs/2008A%26A...487..885O} {487, 885}

\bibitem[\protect\citeauthoryear{{Orienti}, {Murgia}  \& {Dallacasa}}{{Orienti}
  et~al.}{2010}]{2010MNRAS.402.1892O}
{Orienti} M.,  {Murgia} M.,   {Dallacasa} D.,  2010, \mn@doi [\mnras]
  {10.1111/j.1365-2966.2009.16016.x}, \href
  {https://ui.adsabs.harvard.edu/abs/2010MNRAS.402.1892O} {402, 1892}

\bibitem[\protect\citeauthoryear{{Pacholczyk}}{{Pacholczyk}}{1970}]{1970ranp.book.....P}
{Pacholczyk} A.~G.,  1970, {Radio astrophysics. Nonthermal processes in
  galactic and extragalactic sources}

\bibitem[\protect\citeauthoryear{{Parma}, {Murgia}, {de Ruiter}, {Fanti},
  {Mack}  \& {Govoni}}{{Parma} et~al.}{2007}]{2007A&A...470..875P}
{Parma} P.,  {Murgia} M.,  {de Ruiter} H.~R.,  {Fanti} R.,  {Mack} K.~H.,
  {Govoni} F.,  2007, \mn@doi [\aap] {10.1051/0004-6361:20077592}, \href
  {https://ui.adsabs.harvard.edu/abs/2007A&A...470..875P} {470, 875}

\bibitem[\protect\citeauthoryear{{Prakash}, {Licquia}, {Newman}  \&
  {Rao}}{{Prakash} et~al.}{2015}]{2015ApJ...803..105P}
{Prakash} A.,  {Licquia} T.~C.,  {Newman} J.~A.,   {Rao} S.~M.,  2015, \mn@doi
  [\apj] {10.1088/0004-637X/803/2/105}, \href
  {https://ui.adsabs.harvard.edu/abs/2015ApJ...803..105P} {803, 105}

\bibitem[\protect\citeauthoryear{{Rodr{\'\i}guez Zaur{\'\i}n}, {Tadhunter},
  {Rose}  \& {Holt}}{{Rodr{\'\i}guez Zaur{\'\i}n}
  et~al.}{2013}]{2013MNRAS.432..138R}
{Rodr{\'\i}guez Zaur{\'\i}n} J.,  {Tadhunter} C.~N.,  {Rose} M.,   {Holt} J.,
  2013, \mn@doi [\mnras] {10.1093/mnras/stt423}, \href
  {https://ui.adsabs.harvard.edu/abs/2013MNRAS.432..138R} {432, 138}

\bibitem[\protect\citeauthoryear{{Sanders} \& {Mirabel}}{{Sanders} \&
  {Mirabel}}{1996}]{1996ARA&A..34..749S}
{Sanders} D.~B.,  {Mirabel} I.~F.,  1996, \mn@doi [\araa]
  {10.1146/annurev.astro.34.1.749}, \href
  {http://ads.nao.ac.jp/abs/1996ARA%26A..34..749S} {34, 749}

\bibitem[\protect\citeauthoryear{{Skrutskie} et~al.,}{{Skrutskie}
  et~al.}{2006}]{2006AJ....131.1163S}
{Skrutskie} M.~F.,  et~al., 2006, \mn@doi [\aj] {10.1086/498708}, \href
  {https://ui.adsabs.harvard.edu/abs/2006AJ....131.1163S} {131, 1163}

\bibitem[\protect\citeauthoryear{{Soifer}, {Boehmer}, {Neugebauer}  \& {Sand
  ers}}{{Soifer} et~al.}{1989}]{1989AJ.....98..766S}
{Soifer} B.~T.,  {Boehmer} L.,  {Neugebauer} G.,   {Sand ers} D.~B.,  1989,
  \mn@doi [\aj] {10.1086/115178}, \href
  {https://ui.adsabs.harvard.edu/abs/1989AJ.....98..766S} {98, 766}

\bibitem[\protect\citeauthoryear{{Soifer} et~al.,}{{Soifer}
  et~al.}{2000}]{2000AJ....119..509S}
{Soifer} B.~T.,  et~al., 2000, \mn@doi [\aj] {10.1086/301233}, \href
  {http://ads.nao.ac.jp/abs/2000AJ....119..509S} {119, 509}

\bibitem[\protect\citeauthoryear{{Spoon}, {Moorwood}, {Lutz}, {Tielens},
  {Siebenmorgen}  \& {Keane}}{{Spoon} et~al.}{2004}]{2004A&A...414..873S}
{Spoon} H.~W.~W.,  {Moorwood} A.~F.~M.,  {Lutz} D.,  {Tielens} A.~G.~G.~M.,
  {Siebenmorgen} R.,   {Keane} J.~V.,  2004, \mn@doi [\aap]
  {10.1051/0004-6361:20031656}, \href
  {http://ads.nao.ac.jp/abs/2004A%26A...414..873S} {414, 873}

\bibitem[\protect\citeauthoryear{{Surace}, {Sanders}, {Vacca}, {Veilleux}  \&
  {Mazzarella}}{{Surace} et~al.}{1998}]{1998ApJ...492..116S}
{Surace} J.~A.,  {Sanders} D.~B.,  {Vacca} W.~D.,  {Veilleux} S.,
  {Mazzarella} J.~M.,  1998, \mn@doi [\apj] {10.1086/305028}, \href
  {https://ui.adsabs.harvard.edu/abs/1998ApJ...492..116S} {492, 116}

\bibitem[\protect\citeauthoryear{{Teng}, {Wilson}, {Veilleux}, {Young},
  {Sanders}  \& {Nagar}}{{Teng} et~al.}{2005}]{2005ApJ...633..664T}
{Teng} S.~H.,  {Wilson} A.~S.,  {Veilleux} S.,  {Young} A.~J.,  {Sanders}
  D.~B.,   {Nagar} N.~M.,  2005, \mn@doi [\apj] {10.1086/491595}, \href
  {https://ui.adsabs.harvard.edu/abs/2005ApJ...633..664T} {633, 664}

\bibitem[\protect\citeauthoryear{{Thompson}, {Quataert}  \&
  {Murray}}{{Thompson} et~al.}{2005}]{2005ApJ...630..167T}
{Thompson} T.~A.,  {Quataert} E.,   {Murray} N.,  2005, \mn@doi [\apj]
  {10.1086/431923}, \href {http://ads.nao.ac.jp/abs/2005ApJ...630..167T} {630,
  167}

\bibitem[\protect\citeauthoryear{{Thompson}, {Quataert}, {Waxman}, {Murray}  \&
  {Martin}}{{Thompson} et~al.}{2006}]{2006ApJ...645..186T}
{Thompson} T.~A.,  {Quataert} E.,  {Waxman} E.,  {Murray} N.,   {Martin} C.~L.,
   2006, \mn@doi [\apj] {10.1086/504035}, \href
  {https://ui.adsabs.harvard.edu/abs/2006ApJ...645..186T} {645, 186}

\bibitem[\protect\citeauthoryear{{Torniainen}, {Tornikoski}, {Ter{\"a}sranta},
  {Aller}  \& {Aller}}{{Torniainen} et~al.}{2005}]{2005A&A...435..839T}
{Torniainen} I.,  {Tornikoski} M.,  {Ter{\"a}sranta} H.,  {Aller} M.~F.,
  {Aller} H.~D.,  2005, \mn@doi [\aap] {10.1051/0004-6361:20041886}, \href
  {http://ads.nao.ac.jp/abs/2005A%26A...435..839T} {435, 839}

\bibitem[\protect\citeauthoryear{{V{\"a}is{\"a}nen}, {Rajpaul}, {Zijlstra},
  {Reunanen}  \& {Kotilainen}}{{V{\"a}is{\"a}nen}
  et~al.}{2012}]{2012MNRAS.420.2209V}
{V{\"a}is{\"a}nen} P.,  {Rajpaul} V.,  {Zijlstra} A.~A.,  {Reunanen} J.,
  {Kotilainen} J.,  2012, \mn@doi [\mnras] {10.1111/j.1365-2966.2011.20186.x},
  \href {https://ui.adsabs.harvard.edu/abs/2012MNRAS.420.2209V} {420, 2209}

\bibitem[\protect\citeauthoryear{{Vega}, {Silva}, {Panuzzo}, {Bressan},
  {Granato}  \& {Chavez}}{{Vega} et~al.}{2005}]{2005MNRAS.364.1286V}
{Vega} O.,  {Silva} L.,  {Panuzzo} P.,  {Bressan} A.,  {Granato} G.~L.,
  {Chavez} M.,  2005, \mn@doi [\mnras] {10.1111/j.1365-2966.2005.09678.x},
  \href {https://ui.adsabs.harvard.edu/abs/2005MNRAS.364.1286V} {364, 1286}

\bibitem[\protect\citeauthoryear{{Veilleux} \& {Osterbrock}}{{Veilleux} \&
  {Osterbrock}}{1987}]{1987ApJS...63..295V}
{Veilleux} S.,  {Osterbrock} D.~E.,  1987, \mn@doi [\apjs] {10.1086/191166},
  \href {http://ads.nao.ac.jp/abs/1987ApJS...63..295V} {63, 295}

\bibitem[\protect\citeauthoryear{{Veilleux}, {Kim}  \& {Sanders}}{{Veilleux}
  et~al.}{1999}]{1999ApJ...522..113V}
{Veilleux} S.,  {Kim} D.~C.,   {Sanders} D.~B.,  1999, \mn@doi [\apj]
  {10.1086/307634}, \href
  {https://ui.adsabs.harvard.edu/abs/1999ApJ...522..113V} {522, 113}

\bibitem[\protect\citeauthoryear{{Veilleux}, {Kim}  \& {Sanders}}{{Veilleux}
  et~al.}{2002}]{2002ApJS..143..315V}
{Veilleux} S.,  {Kim} D.~C.,   {Sanders} D.~B.,  2002, \mn@doi [\apjs]
  {10.1086/343844}, \href
  {https://ui.adsabs.harvard.edu/abs/2002ApJS..143..315V} {143, 315}

\bibitem[\protect\citeauthoryear{{Veilleux} et~al.,}{{Veilleux}
  et~al.}{2009}]{2009ApJS..182..628V}
{Veilleux} S.,  et~al., 2009, \mn@doi [\apjs] {10.1088/0067-0049/182/2/628},
  \href {http://ads.nao.ac.jp/abs/2009ApJS..182..628V} {182, 628}

\bibitem[\protect\citeauthoryear{{Voges} et~al.,}{{Voges}
  et~al.}{1999}]{1999A&A...349..389V}
{Voges} W.,  et~al., 1999, \aap, \href
  {https://ui.adsabs.harvard.edu/abs/1999A&A...349..389V} {349, 389}

\bibitem[\protect\citeauthoryear{{Yuan}, {Kewley}  \& {Sanders}}{{Yuan}
  et~al.}{2010}]{2010ApJ...709..884Y}
{Yuan} T.~T.,  {Kewley} L.~J.,   {Sanders} D.~B.,  2010, \mn@doi [\apj]
  {10.1088/0004-637X/709/2/884}, \href
  {https://ui.adsabs.harvard.edu/abs/2010ApJ...709..884Y} {709, 884}

\bibitem[\protect\citeauthoryear{{Yun}, {Reddy}  \& {Condon}}{{Yun}
  et~al.}{2001}]{2001ApJ...554..803Y}
{Yun} M.~S.,  {Reddy} N.~A.,   {Condon} J.~J.,  2001, \mn@doi [\apj]
  {10.1086/323145}, \href
  {https://ui.adsabs.harvard.edu/abs/2001ApJ...554..803Y} {554, 803}

\bibitem[\protect\citeauthoryear{{Zhou}, {Wang}, {Wang}, {Wang}, {Yuan}  \&
  {Lu}}{{Zhou} et~al.}{2006}]{2006ApJ...639..716Z}
{Zhou} H.,  {Wang} T.,  {Wang} H.,  {Wang} J.,  {Yuan} W.,   {Lu} Y.,  2006,
  \mn@doi [\apj] {10.1086/499768}, \href
  {https://ui.adsabs.harvard.edu/abs/2006ApJ...639..716Z} {639, 716}

\bibitem[\protect\citeauthoryear{{van Weeren}, {de Gasperin}, {Akamatsu},
  {Br{\"u}ggen}, {Feretti}, {Kang}, {Stroe}  \& {Zandanel}}{{van Weeren}
  et~al.}{2019}]{2019SSRv..215...16V}
{van Weeren} R.~J.,  {de Gasperin} F.,  {Akamatsu} H.,  {Br{\"u}ggen} M.,
  {Feretti} L.,  {Kang} H.,  {Stroe} A.,   {Zandanel} F.,  2019, \mn@doi [\ssr]
  {10.1007/s11214-019-0584-z}, \href
  {https://ui.adsabs.harvard.edu/abs/2019SSRv..215...16V} {215, 16}

\makeatother
\end{thebibliography}

% Don't change these lines
\bsp	% typesetting comment
\label{lastpage}
\end{document}